\documentclass[aps,prb,reprint,superscriptaddress,nofootinbib]{revtex4-2}
\usepackage{amsmath,amsthm}
\usepackage{amssymb}
\usepackage{verbatim}
\usepackage{mathtools}
\usepackage{dsfont}
\usepackage{xcolor}
\usepackage{algorithm}
\usepackage[noend]{algpseudocode}
\usepackage{graphicx}
\usepackage{subcaption}

\usepackage{nicefrac}

\usepackage{thmtools}
\usepackage{thm-restate}

\usepackage{mathrsfs}
\usepackage{nccmath}
\usepackage{placeins}
\usepackage{physics}
\usepackage{comment}
\usepackage{qcircuit}
\usepackage{pgffor}
\usepackage[colorlinks=true,linkcolor=blue,urlcolor=blue,citecolor=blue]{hyperref}
\usepackage{cleveref}
\usepackage{todonotes}

\newcommand{\bea}{\begin{eqnarray}}
\newcommand{\eea}{\end{eqnarray}}

\graphicspath{ {figures} }

\usepackage{ragged2e}

\usepackage[compatibility=false,%
            format=plain,%
            justification=justified,%
            singlelinecheck=false]{caption}

\makeatletter
\renewcommand\@makecaption[2]{%
  \vskip\abovecaptionskip
  {\small #1.\ }\justifying #2\par
  \vskip\belowcaptionskip}
\makeatother

\usepackage{float}

\begin{document}
\title{
Discovery of energy landscapes
towards optimized quantum transport: Environmental effects and long-range tunneling
}

\author{Maggie Lawrence}
\affiliation{Department of Physics, 60 Saint George St., University of Toronto, Toronto, Ontario,  M5S 1A7, Canada}
\affiliation{Vector Institute, Toronto, Ontario, M5S 1M1, Canada}

\author{Matthew Pocrnic}
\affiliation{Department of Physics, 60 Saint George St., University of Toronto, Toronto, Ontario,  M5S 1A7, Canada}

\author{Erin Fung}
\affiliation{Department of Physics, 60 Saint George St., University of Toronto, Toronto, Ontario,  M5S 1A7, Canada}

\author{Juan Carrasquilla}
\affiliation{Institute for Theoretical Physics, ETH Zürich, Zürich, Switzerland}
\affiliation{Vector Institute, Toronto, Ontario, M5S 1M1, Canada}

\author{Erik M. Gauger}
\affiliation{SUPA, Institute of Photonics and Quantum Sciences, Heriot-Watt University, Edinburgh EH14 4AS, United Kingdom}

\author{Dvira Segal}
\affiliation{Department of Chemistry and Centre for Quantum Information and Quantum Control,
University of Toronto, 80 Saint George St., Toronto, Ontario, M5S 3H6, Canada}
\affiliation{Department of Physics, 60 Saint George St., University of Toronto, Toronto, Ontario,  M5S 1A7, Canada}

\begin{abstract}
Carrier transport in quantum networks is governed by a variety of factors, including network dimensionality and connectivity, on-site energies, couplings between sites and whether they are short- or long-range, leakage processes, and environmental effects. In this work, we identify classes of quasi-one-dimensional chains with energy profiles that optimize carrier transport under such influences.
Specifically, we optimize on-site energies using Optax’s optimistic gradient descent and AdaMax algorithms, enabled by the JAX automatic differentiation framework. Focusing on nonequilibrium steady-state transport, we study the system's behavior under combined unitary and nonunitary (dephasing and dissipative) effects using the Lindblad quantum master equation. 
After validating our optimization scheme on short chains, we extend the study to larger systems where we identify systematic patterns in energy profiles. Our analysis reveals that different types of energy landscapes enhance transport, depending on whether inter-site tunneling couplings in the chain are short- or long-range, the existence of environmental interactions, and the temperature of the environment.
Our classification and insights of optimal energy landscapes offer guidance for designing efficient transport systems for electronic, photovoltaic and quantum communication applications.
\end{abstract}
\maketitle

\section{Introduction}

Quantum transport is ubiquitous in any non-trivial quantum system, and in the age of quantum technology must be harnessed to effectively design new devices.
For example, understanding exciton transfer across networks is fundamental for light harvesting technologies, encompassing both organic solar cells \cite{Creatore2013, Fruchtman2016, DeSio2017, Rouse2019, Cavassilas2020, Hu2021QDTransport} and natural photosynthetic complexes \cite{Dorfman2013, Romero2014, Romero2017, Tomasi2021, Zhao2024PhotosyntheticTransport}.
Similarly, quantum transport of charge carriers through arrays of quantum dots 
\cite{Lei2022ChargeTransportQD, Yang2015QDEfficiency, Hu2021QDTransport, Bush2021, Bahadou2025, ContrerasPulido2017, Wang2007, Abdullah2016, Mathe2022, Chen2013, Eastham2013} 
or nanoscale devices 
\cite{Scharf2024AugerDecay, Eckstein2023TrionMechanism, Birkmeier2022ExcitonDynamics, Zhao2018TMDSpintronics, Wang2018GeP3, Prati2013, Barraza-Lopez2012, Avriller2011, Perebeinos2008CarbonNanotubes} 
plays a critical role in the development of quantum electronic devices.  
Also, the transmission of quantum information, whether of particles or light, across quantum networks is an increasingly active area of research, driven by the growing interest in quantum information processing. Understanding these systems is important for quantum simulations and quantum search algorithms; see, e.g., Ref. \citenum{QWSearch}. 

In the examples listed above, the transfer of particles and excitations is typically not purely quantum coherent: interactions with the surrounding environment, such as phonons in the material \cite{DeSio2017,Liu2019}, electromagnetic fields \cite{Davidson2022}, or other environmental degrees of freedom, introduce dephasing and dissipation effects. The resultant transfer behavior is thus defined by the non-trivial confluence of coherent and incoherent effects. In practice, the transport of particles under environmental influences is often modeled using quantum walks \cite{Rani_Dutta_Banerjee_2024, Gao2024, loebens2024quantumwalks, konno2023crossover} or with a master equation \cite{Coates2021, Coates2023, DavidsonPollockGauger2021, Davidson2022, Schinabeck2016, Schinabeck2020, he2023transport, Kraft2025}. In this context, ``quantum walks" \cite{QWRev}, the analogue of classical random walks, describe the propagation of quantum particles on networks.

Optimizing transport processes in open quantum systems (OQS) remains an open and actively-investigated question. 
In this work,  our goal is to use modern optimization techniques to discover energetic landscapes that promote transport {\it without imposing a priori} structural assumptions. Our analysis focuses on one-dimensional chains, exploring both short-range and long-range tunneling regimes, and considering the effects of dephasing and thermal relaxation.

A large body of research has shown that in certain cases, for example within disordered systems, interactions with the environment leading to dynamical noise can {\it enhance} quantum transport and define an optimal region of operation. This phenomenon is known as Environmental Noise-Assisted Quantum Transport (ENAQT). While first investigated theoretically \cite{Alan08,Plenio08,Plenio09, Plenio10, Plenio12,Cao13,Plenio21,mohseni2014energy, trautmann2018,zerah-harush2018, zerah-harush2020, cygorek2022, kurt2023}, this effect was recently demonstrated in chains of trapped ions \cite{maier2019environment}, and in the solid-state within nanocrystal superlattices \cite{Blach25}.  In fact, related turnover trends of reaction rates and electron flux with environmental interactions in the form of friction coefficients or decoherence rates are well known, and these effects have been investigated in both classical and quantum systems \cite{Kramer50,Davis97,Segal00}.

In addition to environmental effects, the connectivity of the quantum network and its energy landscape play a crucial role in determining the efficiency of transport. Depending on these factors, the system may exhibit localized, ballistic, or various forms of diffusive transport dynamics. Localization refers to suppression of the propagation of particles, where carriers remain confined near their initial site rather than traversing the network. The specific form of localization depends on the system: Anderson localization arises from disorder in site energies or tunnelings \cite{anderson1958absence,  evers2008_anderson_transitions,lagendijk2009}, while Wannier-Stark localization occurs in systems with a linear, ramp-like energy level gradient \cite{wannier1962dynamics, schulz2019stark, Jacob2024DephasingTransport}. Localization can also occur due to many body interactions \cite{Abanin19}.

In many applications, such as solar energy harvesting, localization is undesirable as it hinders efficient transfer of excitations from their point of generation to the collection site. Promoting delocalization through coherent or noise-assisted transport is therefore a key design goal in such systems \cite{galindo2015dendrimer, zhang2015photosynthesis, resendiz2021noise, rathbone2018coherent}.

A substantial body of research is focused on analyzing {\it particular} system designs, examining how a specific structure may lead to localization \cite{mostarda2013, fischer2016dynamics,EvangelineRebecca2019, Coates2021, Demuth2022, Jen2022}, or how particular systems respond to different environmental conditions, including optimization of the surrounding environments towards maximal efficiency \cite{Alan08,Cao09,Kassal09,mohseni2014energy, trautmann2018,Liu2019,zerah-harush2018,zerah-harush2020,Plenio21,Tomasi2021, Jacob2024DephasingTransport, Langlett2023Fermionic, sarkar2020environment}. 
In contrast, relatively little effort \cite{ Lorber2024QCN, Davidson00} has been devoted to the {\it discovery} of system configurations or energy landscapes that promote efficient transport {\it without presupposing structural motifs}, despite the identification of the energetic landscape as the primary factor for  transport efficiency \cite{DavidsonPollockGauger2021}.

In this paper, our goal is to discover structures
that exhibit enhanced quantum transport under the combined influence of coherent and dephasing or dissipative environments, and to develop intuition into the advantage of these structures.
Specifically, we focus on quasi one-dimensional tight-binding chains in which inter-site tunneling couplings are short- or long-range
and optimize the chains' energy landscape towards transfer of carriers from an entry site to a designated collection site. Environmental effects in the form of dephasing or thermally-induced transitions are incorporated by adding different noise terms using the Lindblad quantum master equation formalism. 
While prior studies employed a {\it forward} approach, fixing the system’s energy structure and then analyzing the resulting behavior, as was done in studies of photosynthetic proteins \cite{Alan08, Plenio08, Plenio09, Plenio10}, we adopt an {\it inverse} method, searching for system energy configurations that optimize transport efficiency.

\begin{figure}[t!]
\includegraphics[width=\linewidth]{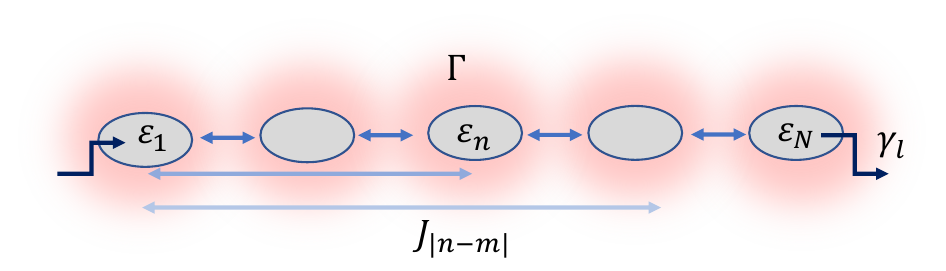}
\caption{Scheme of the model: we study transport through a chain of $N$ sites, with tunneling elements $J_{|n-m|}$ between sites $n$ and $m$. Carriers enter site 1 and leave at site $N$ at a rate $\gamma_l$. Local dephasing is enacted on each site (with a rate constant $\Gamma$ in OQS Model I), indicated here with red semitransparent circle on each site. Given tunneling elements, leakage rate, and the dephasing rate, we search for the set $\varepsilon_{1}$ to $\varepsilon_N$ that maximizes the population flux from site $N$.}
\label{fig:figure1}
\end{figure}

Figure \ref{fig:figure1} illustrates our setup: a chain of $N$ sites, with possibly long-range tunneling coupling elements between sites, leakage rate at the collection site, and environmental effects. We search for the sites' energy profile that maximizes the outgoing flux.

Assuming a fixed environmental influence and network connectivity, whether restricted to short-range or including long-range couplings, we employ gradient ascent, performed with automatic differentiation, to optimize the on-site energies of local sites toward high population transfer.
As we show in this study, the resulting energy landscapes reveal distinct classes of optimal configurations, which vary depending on the range of tunneling, the magnitude and nature of the environmental noise (dephasing or dissipative), and the system's temperature.  
Our work offers guidelines for the type of configuration that optimizes transfer efficiency. Moreover, we gain insights into what makes certain structures optimal and what the underlying transport mechanisms are in different parameter regimes.

The structure of this paper is as follows:
in Section \ref{sec:oqs} we discuss the model of interest and the Lindblad Quantum Master Equation (QME). 
We further describe our figure of merit, the population flux, and the gradient-ascent algorithms used to optimize the energy landscape of the system. Appendix \ref{AppendixA} provides additional details on the optimization process.
To validate our protocol and provide further insights for results, in Sec. \ref{sec:3siteoptim} we focus on 3-site models, which can be solved analytically under some simplifications. 
Section \ref{sec:manysiteoptim} is devoted to the discovery of optimal structures for longer chains using our optimization process, with more examples presented in Appendix \ref{AppendixB}. We conclude in Section \ref{sec:summary}.

\section{Model and Dynamics}
 \label{sec:oqs}

We focus on quantum transport in quasi-one-dimensional systems that are coupled to local environments (see Fig. \ref{fig:figure1}). In order to focus on the essence of the problem, we make it more tractable by introducing several simplifying assumptions. 
First, we consider the presence of only a single carrier within the quantum system, physically corresponding to a single charge or excitation. 
Second, we include the environment in a phenomenological way, using Lindblad dissipators \cite{breuer2007_decoherence}. The advantage of this description is that it applies to different environments, such as intra- or intermolecular vibrations \cite{sarkar2020environment, juhasz2018vibrations}, phonons \cite{Eastham2013, Jager2022}, photons \cite{giri2025entanglement} or electrons \cite{Rosati2014}.
We refer to our models as ``quasi-one-dimensional" since we test both short-range and long-range tunneling coupling scenarios, where in the latter case there is essentially an all-to-all connectivity in the system. In what follows, we work with units of $\hbar\equiv1$ and $k_B\equiv1$.

In the site-local basis, the Hamiltonian of the chain is given by a tight-binding model,
\begin{equation} \label{ham}
    \hat H_S = \sum_{n=1}^{N} \varepsilon_n \ket{n} \bra{n} + \sum_{n \neq m} J_{|n-m|} \ket{n} \bra{m}.
\end{equation}
Here, \(\varepsilon_n\) is the energy of site \textit{n}, and \(J_{|n-m|}\) is the tunneling energy between site \textit{m} to site \textit{n}. 
For simplicity, these tunneling elements are chosen to be real and positive. 
We use a power-law function to describe the tunneling energies,
\begin{equation} \label{powerlaw}
    J_{|n-m|} = \frac{J_{max}}{|n-m|^\alpha},
\end{equation}
with $J_{max}$ as the magnitude of the nearest-neighbors tunneling coupling. In simulations, we use $\alpha=1$ to represent long-range coupling, while $\alpha=3$ supports shorter-range tunnelings. Throughout this work, we distinguish between two tunneling models: the long-range model, characterized by significant tunneling beyond nearest neighbors, and the short-range model, where tunneling is effectively limited to nearest neighbors.

The carriers interact with environments --- typically phonons, 
local impurities, 
and other carriers. 
Microscopically, an environment can be captured in the harmonic approximation by a collection of harmonic modes of frequency $\omega_k$,
\begin{equation} \label{hamb}
   \hat H_B = \sum_{k} \omega_k \hat b^\dagger_k \hat b_k.
\end{equation}
Here, \(\hat b^\dagger_k\) and \(\hat b_k\) are the creation and annihilation operators for the boson mode of the bath. The interaction between the system and bath is typically given in a bipartite form,
\begin{equation} \label{hami}
    \hat H_I = \sum_{n=1}^N \hat L_n \sum_k g_{nk} \otimes \left(\hat b_k^{\dagger}+\hat b_k\right).
\end{equation}
Here, $g_{nk}$, taken to be real-valued without loss of generality, denotes the coupling energy between the $k^\text{th}$ bath mode and a local system operator $\hat L_n$ acting on site $n$.

We consider two models for system-bath couplings and specify the operators of the system that couple to the baths, $\hat L_n$. The first case, OQS Model I, concerns local dephasing effects, as described by trapped-ion experiments \cite{maier2019environment}.
In OQS Model II, a finite temperature environment couples to a local charge density, a scenario relevant for electron-phonon coupled systems \cite{Fruchtman2016}.

\subsection{OQS Model I}
\label{sec:OQSI}

We assume that the environment acts locally and in an uncorrelated manner on each site to dephase the state. The local Lindblad QME is given by
\begin{equation} 
\label{lindpd}
    \dot{\rho} = -i\left[\hat H_S, \rho\right] + \sum_{n=1}^N \Gamma_n \left(\hat L_n \rho \hat L_n^\dagger - \frac{1}{2} \left\{\hat L_n^\dagger \hat L_n, \rho\right\} \right).
\end{equation}
Here, the Lindblad jump operators are given by $ \hat L_n = \ket{n} \bra{n}$, $\Gamma_n$ are the dephasing rate constants, and the dynamics evolve in the site-local basis. 
In this case, the microscopic information from Eqs. \eqref{hamb} and \eqref{hami} is not explicitly used, but one can associate $\Gamma_n$ with a microscopic model of the bath \cite{Coates2021}. It should be noted that in the energy basis, the local dephasing model corresponds to an unstructured infinite temperature bath. This is because, for asymptotically small $\gamma_l$,  the steady state solution of the dynamics  corresponds to the completely mixed state with equal populations at each site, independent of energy profile \cite{Coates2021,Turkeshi2021}. A nonzero value for $\gamma_l$ affects the population distribution due to the nonequilibrium setting, thus showing deviations from the maximally mixed state. However, it can be readily proved that the nonequilibrium distribution is nonthermal, even if $\Gamma_n$ is made site- and temperature-dependent.

We assume that all sites suffer a comparable dephasing 
and set $\Gamma_{n}\equiv\Gamma$. Theory, simulations, and experiments \cite{Scholes14,maier2019environment,Blach25} indicate that noise-assisted quantum transport is most pronounced when $\Gamma/J_{max}\sim 1$. 
Accordingly, we focus on this regime and choose parameters of comparable magnitude.

We comment that it is common to employ the Lindblad QME in the site basis and with local dephasing effects in order to capture local dynamical noise. Examples include recent studies on relaxation timescales in Lindbladian dynamics \cite{times1,times2}, transport in networks \cite{Prosen23}, and quantum simulations on quantum hardware \cite{Plenio23}. 
In such studies, the dephasing rates are taken over a broad range of values compared to the system energy scale. 

\subsection{OQS Model II}

In OQS Model II, the environment still acts locally and independently on each site, but we consider a finite-temperature bath and solve the problem in the energy basis of the system; that is, we employ the global Lindblad QME (also referred to as the exciton-basis QME in the context of exciton transport). In the global basis, transition rates between different eigenstates of the system depend
on the temperature of the environment and the energy difference between those states. 
We write the global Lindblad QME as
\begin{equation} \label{lindT}
    \dot{\rho} = -i \left[\hat H_S, \rho\right] + \sum_{a \neq b} W_{ab} \left(\hat F_{ab} \rho \hat F_{ab}^\dagger - \frac{1}{2} \left\{\hat F_{ab}^\dagger \hat F_{ab}, \rho \right\} \right).
\end{equation}
Here, the \(\hat F_{ab}\) jump operators describe transitions between eigenstates \textit{a} and \textit{b} of the system's Hamiltonian, i.e. $\hat F_{ab} = \ket{\phi_a} \bra{\phi_b}$,
where \(\ket{\phi_a}\) is the \(a^{th}\) eigenstate of \(\hat H_S\), with eigenenergy \(\omega_a\).
The transition rates \(W_{ab}\) in Eq. \eqref{lindT} are given by 
\begin{equation} \label{Wab}
    W_{ab} = \sum_{n=1}^N S_{ba} \bra{\phi_a} \hat L_n \ket{\phi_b} \bra{\phi_b} \hat L_n \ket{\phi_a},
\end{equation}
where \(\hat L_n=|n\rangle\langle n|\) are the site-local operators and
$S_{ab}$ corresponds to the bath spectral function \cite{Davidson2022},
calculated at the specific transition,
\begin{equation} \label{Sab}
    S_{ab} = \Gamma_0 \left|\omega_{ab}\right| \left(n_{BE}\left(|\omega_{ab}|\right) + \Theta\left(\omega_{ab}\right) \right).
\end{equation}
Here, we assume that the baths are characterized by ohmic spectral functions, but one can readily select other forms.
The above factor depends on the temperature of the bath;
\(n_{BE}(\omega)\) is the Bose-Einstein occupation factor $ n_{BE} (\omega) = \frac{1}{e^{\omega/T} - 1}$. Other parameters are
\(\Gamma_0\) as a dimensionless constant that dictates
the strength of system-bath coupling, \(\omega_{ab} = \omega_a - \omega_b\), and 
\(\Theta(\omega)\) as the Heaviside step function.
Eq. \eqref{Sab} imposes the detailed balance relation, thus ensuring that transitions to lower energy states are more probable than transitions to higher-energy state; 
if \(\omega_{ab} > 0\), then  $S_{ab} = \Gamma_0 \omega_{ab} \frac{e^{\omega_{ab}/T}}{e^{\omega_{ab}/T} - 1}$,   $ S_{ba} = \Gamma_0 \omega_{ab} \frac{1}{e^{\omega_{ab}/T} - 1} $.
The dynamics imposed by Eq. (\ref{lindT}) decouples the population and coherences by using the so-called secular approximation. It relies on the assumption that energy levels are sufficiently separated from each other such that internal coherent dynamics have timescales shorter than the overall excitation decay time. 


\subsection{Nonequilibrium steady state and measures for flux} \label{sec:SSsolve}
 
In the literature, different measures have been suggested to quantify the ability of a network or a conductor to efficiently transfer carriers from an initial site to a target point. This includes the mean first passage time and other transient measures \cite{Alan08,Kassal09,maier2019environment, ueno2025distributingentanglementquantumspeed, Zhang2024SteadyTransport}, as well as steady-state (or quasi-steady-state) measures, such as a rate constant and flux \cite{zerah-harush2018, zerah-harush2020, Coates2021, Kalantar19, Dutta2020CoherenceSteadyStates, Dutta2021OutOfEquilibrium, Coates2023}. 
Here, we set the system under nonequilibrium steady-state conditions, and our measure for transfer is the population flux out of site $N$.

The nonequilibrium steady state is constructed by enforcing a constant flux of carriers entering site $1$ and leaving site $N$. These sites were chosen with the aim of achieving transport over the longest distance allowed given a specified number of sites. Different combinations of injection and leakage sites may lead to different transport mechanisms being observed, see Refs. \cite{maier2019environment, Dutta2021OutOfEquilibrium}. We consider both Eqs. \eqref{lindpd} and \eqref{lindT}, and formally write them as \(\dot{\vec{\rho}} = M \vec{\rho} \), with the density matrix written in vectorized form following the methods described in Ref. \cite{LandiRevModPhys2022}.

To enforce the nonequilibrium steady state, first, we add the following jump operator to the Lindblad QME (written here in the site basis),
%
    $\hat L_{l} = \ket{1} \bra{N}$,
%
along with a temperature-independent rate constant, denoted by \(\gamma_l\).  This term corresponds to a leak and injection (or trapping) process. Second, in the steady state, we have \(\dot{\rho} = 0 \);  we find the solution to this equation, $\rho_{SS}$, by replacing one of the dependent rows in $M$ (in our case, the first row) with the population normalization condition and solving $\tilde M\vec{\rho}^{SS} = \vec{u}$, where $\vec{u}$ is a vector of zeros besides the normalization value in the first row. $\tilde M$ is the modified Liouvillian matrix after adding the leakage jump operator and the normalization equation. This allows us to obtain the steady state as a matrix inversion task, rather than by obtaining the eigenvectors of $M$, which presents difficulties with automatic differentiation (see Appendix \ref{AppendixA}).

Our measure of transfer is the nonequilibrium steady-state flux from the exit site $N$ of the system. It is given by
\begin{equation} 
\label{eta} \eta = \gamma_l \bra{N} \rho^{SS} \ket{N}.
\end{equation}
Maximizing $\eta$ directly implies that one tries to maximize the population at the exit site. It is upper-bounded by $\eta\leq \gamma_l$.

The flux measures the rate of population leakage from the exit site $N$. Since the jump operators that we use to describe the environment do not deplete population, the flux is constant along the chain, though for nonlocal tunneling, it has to be calculated along all contributing bonds \cite{Liu2019}.

We now make several comments on the nonequilibrium steady state transport framework, the measure we use, and its relation to other measures.

First, we know that when $\gamma_l=0$, and for a nonzero $\Gamma$, the system reaches in the long time limit its equilibrium steady state solution, which for OQS Model I is the completely mixed state. For $\gamma_l\neq 0$, the fixed point is a nonequilibrium steady state.

Second, several studies have employed transient measures to quantify transfer efficiency. In particular, Ref. \cite{maier2019environment} used the following definition, $\tau_T=\frac{1}{T}\int_0^{T} p_N(\tau) d\tau$, with $p_N(t)$ as the time-evolving population of the target site, $N$. The integration time $T$ must be chosen carefully: it has to be chosen such that time evolution is long enough to build population in the target site from the initial condition, and short enough to be distinct from the final equilibrium state. For a chain with population initialized at site 1 with the target site $N$, a common choice is $T=N/J_{max}$, where $J_{max}$ is the nearest neighbor tunneling energy and $N$ the site-to-site distance between initial to target sites \cite{maier2019environment}. 
However, this choice of integration time can significantly affect results. Moreover, the expression for $T$ must generally be re-evaluated and adjusted for different energy profiles and dephasing strengths. Overall, time-dependent measures of this kind may lack robustness, and varying  $T$ influences the inferred transport efficiency.

Third, we note that different measures correspond to different experimental settings. Experiments may be performed as an initial-condition problem, where an excitation is prepared at an entry site and the transfer efficiency is quantified by the rate or probability of arrival at a designated target site. In contrast, the nonequilibrium steady-state framework corresponds to a boundary-condition problem, analogous to a current–voltage setup, where a constant drive or input maintains a steady current across the system.
These two scenarios, imposing initial conditions versus enforcing nonequilibrium boundary conditions, can generally lead to different trends. This does not imply that either approach is incorrect; rather, it highlights that they describe distinct experimental settings. In Ref. \cite{Kalantar19}, we compared transient and nonequilibrium steady state measures and demonstrated cases where the two approaches yield consistent results, as well as situations in which they diverge.

\subsection{Optimization}
\label{sec:optimization}

At the heart of this work lies a fundamental question: which classes of energy profiles, $\vec{\varepsilon}=(\varepsilon_1,\varepsilon_2,...,\varepsilon_N)$,
maximize transport flux? We focus in particular on how long-range versus short-range tunneling shapes this optimization, and how environmental coupling and finite-temperature effects further influence the optimal energy landscape. Because the dynamics entwine both coherent and incoherent processes, the resulting optimal profiles are expected to emerge from the interplay between these competing mechanisms.

The maximized population flux is obtained using one of two optimization algorithms: optimistic gradient ascent (OGA) or AdaMax. In both cases, an initial set of on-site energies are iteratively updated toward improving performance. 

For OGA \cite{mokhtari2019unifiedanalysisextragradientoptimistic}, in step $k$, site energies are updated according to
\begin{equation} \label{oga}
	\vec{\varepsilon}_{k+1} = \vec{\varepsilon}_k + 2h \nabla_\varepsilon \eta\left(\vec{\varepsilon}_k\right) - h\nabla_\varepsilon \eta \left(\vec{\varepsilon}_{k-1}\right).
\end{equation}
In AdaMax \cite{deepmind2020jax, kingma2017adammethodstochasticoptimization}, site energies are updated according to
\begin{align} \label{adamax}
    \vec{\varepsilon}_{k+1} &= \vec{\varepsilon_{k}} + h m_k/(v_k(1-\beta_1^k)) \\
    m_k &= \beta_1 m_{k-1} + (1-\beta_1) \nabla_\varepsilon \eta(\vec{\varepsilon}_k) \nonumber \\
    v_k &= \textrm{max}(|\nabla_\varepsilon \eta(\vec{\varepsilon}_k) + \tilde{\epsilon}|, \beta_2 v_{k-1}). \nonumber
\end{align}
Here, \(\vec{\varepsilon}_k\) is a vector consisting of the values of site energies after $k$ iterations, \(h\) is the learning rate, $m_k$ is an exponentially-decaying moving average of the gradient, $v_k$ is an exponentially-decaying moving average of the square of the gradient, $\beta_1$ is the decay rate of $m_k$, $\beta_2$ is the decay rate of $v_k$, and $\tilde{\epsilon}$ is a small constant used to prevent division by zero. The energy profile \(\vec{\varepsilon}_{max}\) that 
maximizes the flux in equation \eqref{eta} is the zero of the gradient taken with respect to the system energies,
\begin{equation} \label{maxeta}
    \eta\left(\vec{\varepsilon}_{max}\right) = \max(\eta) \Longrightarrow \nabla_{\varepsilon} \eta \left(\vec{\varepsilon}_{max}\right) = \vec{0}.
\end{equation}
Eq. \eqref{maxeta} provides a stopping condition for the gradient ascent algorithm: 
exit the loop if \(\left|\nabla_\varepsilon \eta\left(\vec{\varepsilon}_k\right)\right| < \epsilon \), 
where \(\epsilon\) is some small tolerance. No constraints were placed on the energies -- instead, a maximum number of steps was defined to stop diverging optimizations from iterating indefinitely.
Python libraries JAX and Optax \cite{deepmind2020jax} are used to calculate the flux derivatives and perform the OGA and AdaMax optimization.

Gradient-based optimization algorithms are sensitive to their starting condition. We also expect population flux as a function of site energies to have several local optima. To cover the parameter space thoroughly and identify many maxima of interest, 
we perform optimization runs starting from many initial energy profiles.
Since the system's behavior depends only on energy differences and not on the absolute value of energy, the energy of the first site is fixed to zero. 
Then, a ``hypergrid" is constructed over the remaining energies \((\varepsilon_2, ..., \varepsilon_N)\) and the initial conditions are randomly sampled from this grid.

For chains with only a few sites, an exhaustive search over the full hypergrid is feasible. However, as system size grows, this approach quickly becomes computationally prohibitive, and, in practice, unnecessary. In larger systems, efficient exploration of the energy landscape can be achieved through random initializations
followed by gradient ascent optimization.

In Appendix \ref{AppendixA}, we benchmark several optimization algorithms on a three-level system. Among them, OGA and AdaMax consistently performed best for our class of problems, which often feature a characteristic ``ridge" of near-optimal solutions.

\subsection{Discussion over transport mechanisms and choice of parameters}
\label{sec:transport}

Depending on the range of parameters, in the coherent case one can observe sequential tunneling, deep tunneling, or ballistic motion. Furthermore, long-range tunneling coupling terms, such as $J_2$, lead to interference effects with, e.g., short-range processes emerging from tunneling terms $J_1$. Such interference effects can be constructive or destructive (see Section \ref{sec:3siteoptim}).
When environmental effects are included, they can enhance the flux by destroying the Anderson localization effect through level broadening, or suppress ballistic transport.
Strong environmental effects in the form of local dephasing are known to freeze dynamics, interpreted as the quantum Zeno effect \cite{Kassal09,zerah-harush2018, maier2019environment}. Furthermore, the finite-temperature dissipation model (OQS II) can enact diffusion-like transport \cite{Lee2015}.

We now provide a more organized discussion of transport regimes in quasi one-dimensional systems. For a detailed discussion, we refer readers to reviews such as \cite{Dhar08,LandiRevModPhys2022}:

(i) Quantum coherent transport:
when the dephasing rate is zero or much smaller than the tunneling amplitudes ($\Gamma_n \ll J_{|n-m|}$), the dynamics and steady state are governed by the unitary (coherent) part of the evolution. We therefore refer to this regime as coherent transport.
Within this regime, several distinct scenarios can be identified depending on the system’s energy landscape.
If the terminal sites lie significantly lower or higher in energy than the intermediate sites, deep tunneling processes occur, characterized by an exponentially decaying flux with increasing system length.
In contrast, for a flat energy profile, ballistic transport arises, where transport measures such as flux, current, or efficiency become independent of system size.
Disordered systems, on the other hand, are expected to exhibit localization in the thermodynamic limit.
Since we do not analyze the scaling of transport properties with system length in this work, we infer possible underlying transport mechanisms directly from steady-state features, the coherences and population patterns.
A uniform (flat) population profile, for example, is indicative of ballistic motion.


(ii) Incoherent regime:
when the dephasing or dissipation rate dominates all other relevant energy scales, i.e., the tunneling elements, site energies, and leakage rates ($\Gamma \gg J_{|n-m|}, \varepsilon_n, \gamma_l$), transport becomes diffusive with system length. In this regime, transport measures (e.g., flux or current) are expected to scale as $1/N$. 
Since we do not explicitly analyze the length dependence here, we instead use Fick’s law, manifested as a constant population gradient, as an indicator of diffusive motion.
Furthermore, the dependence of the flux on $\Gamma$ manifests a Kramer's turning behavior from $\eta\propto\Gamma$
in the weak dissipative limit to $\eta\propto1/\Gamma$ under strong dissipation \cite{Cao13}.


(iii) Intermediate regime:
when the dephasing rate $\Gamma$ and the dissipation rate are of a comparable magnitude to the tunneling amplitudes, $J_{|n-m|}$, the dynamics and resulting steady state reflect a competition between coherent and incoherent processes. In this regime, neither $\Gamma$ nor the tunneling terms can be neglected (see, for example, Eq.~(\ref{eq:etaG})), and the resulting flux arises from the influence of both effects.
This interplay can, in some cases, enhance transport efficiency, particularly in energetically disordered systems, a phenomenon known as environmental noise-assisted quantum transport.

In this study, we select parameters for the chains such that competing transport effects contribute to the population flux. That is, we set $J_{max}$, $\gamma_l$, the dephasing rate constant $\Gamma$ and the temperature $T$ to have similar values, thus achieving transport where they all participate. The purpose of the study was not to identify dominant transport mechanisms; this discussion is meant to highlight different possible underlying transport mechanisms that dominate in different regimes.

With our choice of \(\hbar = k_B = 1\), every physical parameter that we use effectively has units of energy and may be compared directly to the site energies that we find through optimization. This allows us to easily comment on the site energy scale that plays best with our chosen tunneling and environmental effects. 
To convert these values into physical units, one would divide the tunneling elements or dephasing rates by $\hbar$ to obtain parameters in units of inverse time, where $\hbar$ is the reduced Planck's constant given in the same energy units as the site energies. For example, a typical energy scale may be on the order of meV \cite{Creatore2013, Blach25}, in which case \( \hbar = 6.58 \cdot 10^{-13} \text{meV} \cdot \text{s}\). The temperature in degrees K is obtained similarly by dividing by $k_B$, the Boltzmann's constant expressed in the appropriate energy unit.

As for the magnitude of the population flux, given the definition, Eq. (\ref{eta}), for an $N$-site chain we anticipate it to be about $\eta \approx \gamma_l/N$ in the steady state limit (assuming for simplicity equal population). As such, for our choice of parameters and for $N=10$ sites, we get $\eta =0.01$, which translates to $\eta \approx$ 16 ps$^{-1}$
if we consider parameters in units of eV.

\begin{figure*}
  \includegraphics[width=\linewidth]{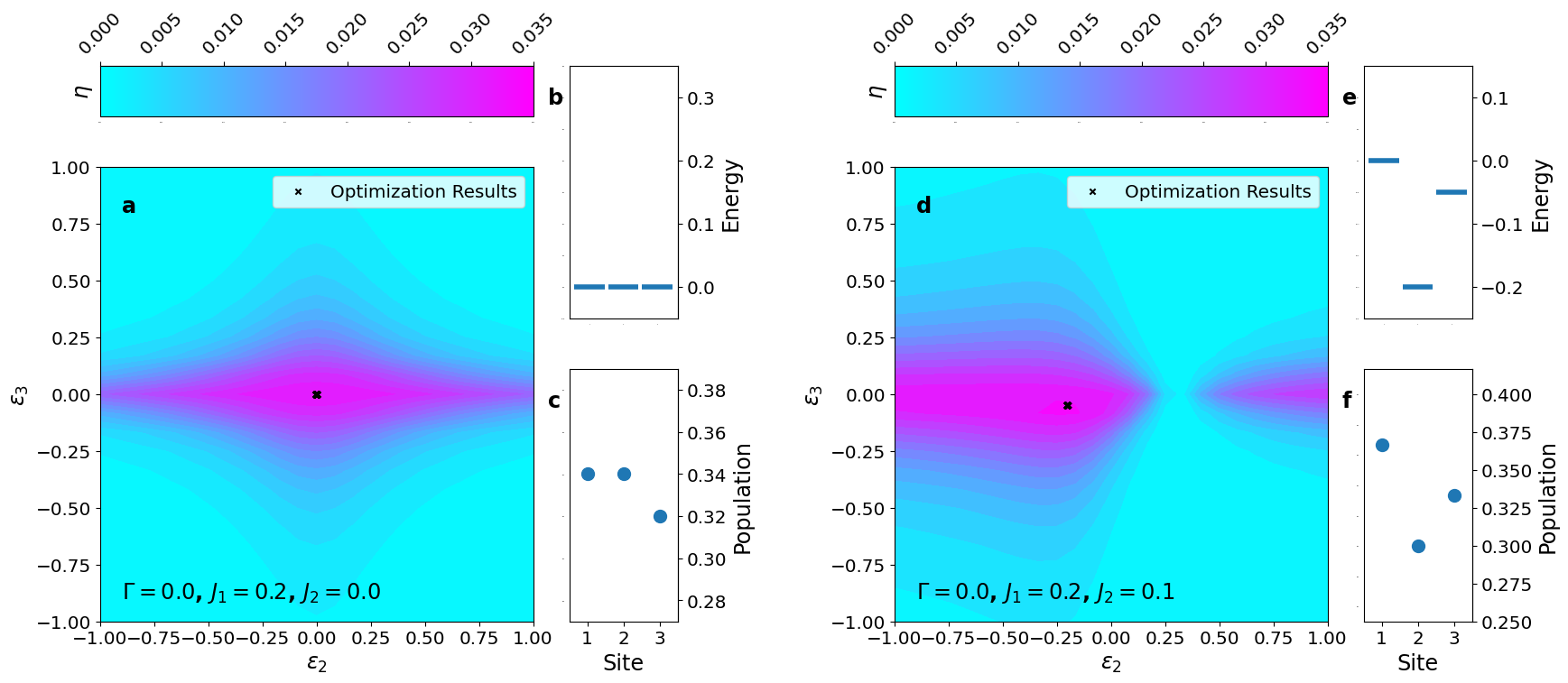}
\caption{Coherent Model: population flux in the absence of environmental effects.
Left, (a)-(c): 3-site model with nearest-neighbor tunneling, $J_1=0.2$ and $J_2=0$.
(a) Flux map as a function of $\varepsilon_2$ and $\varepsilon_3$, fixing $\varepsilon_1=0$.
(b) Level diagram of the optimal configuration,  \(\varepsilon_2 \in [-3.0, 3.0] \cdot 10^{-4}\)
(these values yield the same flux) and \(\varepsilon_3 = 1.6 \cdot 10^{-11}\), 
with the optimal flux \(\eta = 0.032\).
(c) Steady-state population under the optimal profile, \(\rho_{11}^{SS} = \rho_{22}^{SS} = 0.34\), \(\rho_{33}^{SS} = 0.32\).
Right, (d)-(e): 3-site model with beyond-nearest-neighbor tunneling, \(J_1 = 0.2\) and \(J_2 = 0.1\).
(d) Population flux map.
(e) Level diagram of the optimal configuration,  \(\varepsilon_2 = -0.200\) and \(\varepsilon_3 = -0.050\)
leading to \(\eta = 0.033\).
(f) Steady-state population under the optimal profile, \(\rho_{11}^{SS} = 0.37\), \(\rho_{22}^{SS} = 0.30\), \(\rho_{33}^{SS} = 0.33\). 
Other parameters are  \(\Gamma = 0\), leakage rate constant \(\gamma_l = 0.1\).
In panels (a) and (d), results of the OGA algorithm are marked by a black x.}
\label{fig:figure2}
\end{figure*}
\begin{figure*}
\includegraphics[width=\linewidth]{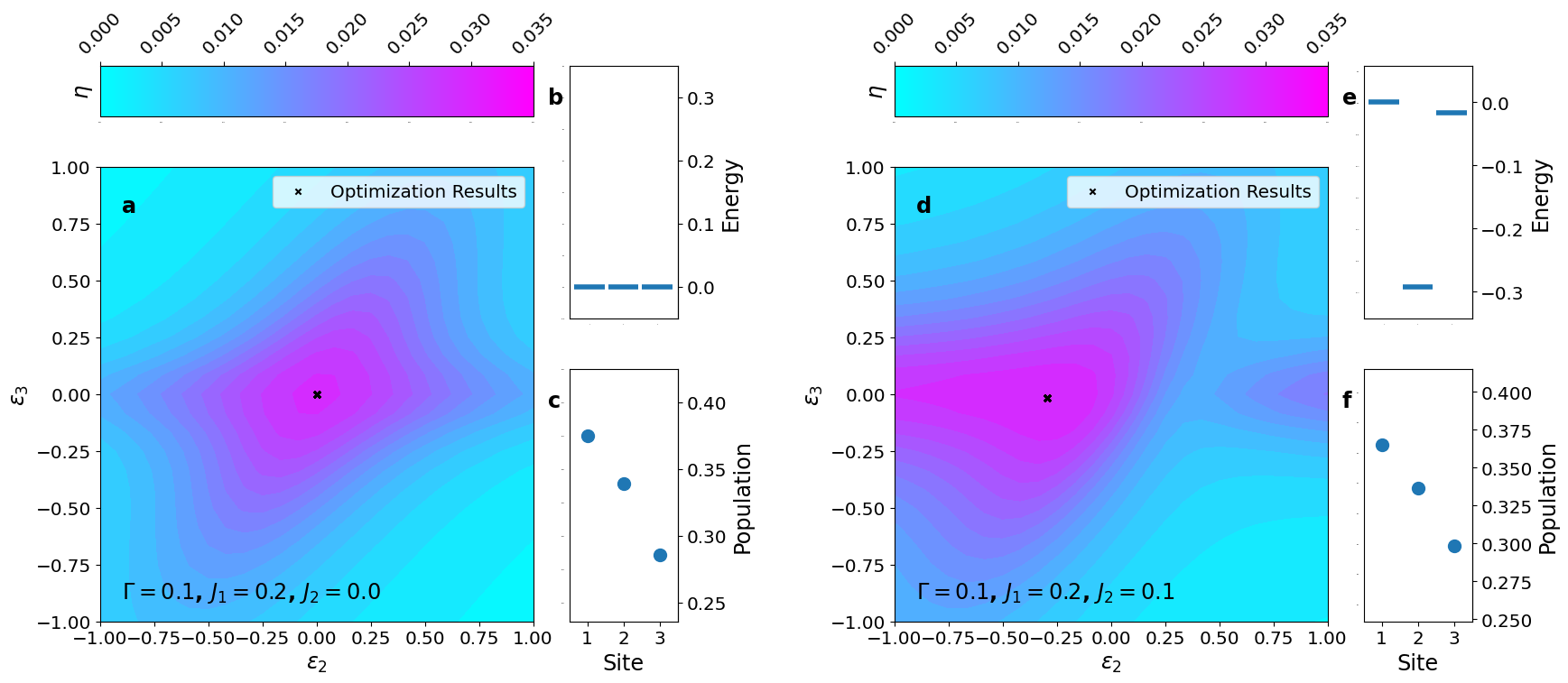}
\caption{OQS Model I: population flux in the presence of environmental effects with $\Gamma=0.1$.
Left, (a)-(c): model with nearest-neighbor tunneling, $J_1 =0.2$ and $J_2=0$. 
(a) Population flux map. (b) Level diagram of the optimal configuration,
$\varepsilon_2= 1.3 \cdot 10^{-5}$, $\varepsilon_3= 5.8 \cdot 10^{-6}$, resulting in $\eta=0.029$.
(c) Steady-state population under the optimal profile, \(\rho_{11}^{SS} = 0.37\), \(\rho_{22}^{SS} = 0.34\), \(\rho_{33}^{SS} = 0.29\).
Right, (d)-(e): 3-site model with next-nearest-neighbor tunneling, \(J_1 = 0.2\) and \(J_2 = 0.1\).
(d) Population flux map.
(e) Level diagram of the optimal configuration,  $\varepsilon_2= -0.292$, $\varepsilon_3= -0.017$, leading to
$\eta=0.030$.
(f) Steady-state population under the optimal profile, \(\rho_{11}^{SS} = 0.36\), \(\rho_{22}^{SS} = 0.34\), \(\rho_{33}^{SS} = 0.30\).
We set \(\gamma_l = 0.1\).
In both (a) and (d), optimization results of the OGA algorithm are marked by a black x.}
\label{fig:figure3}
\end{figure*}

\begin{figure*}[htbp]
    \centering
    \includegraphics[width=\linewidth]{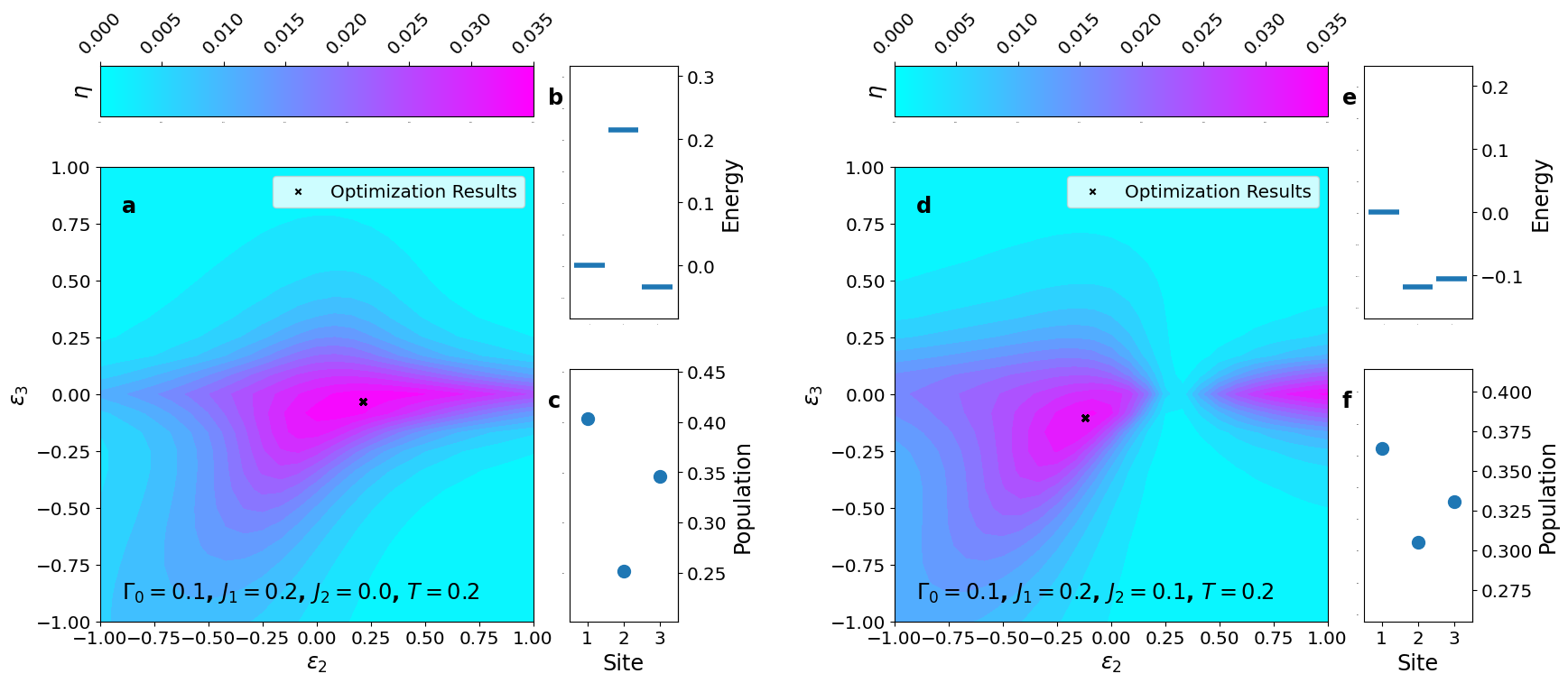}
 \caption{OQS Model II: population flux in the presence of environmental effects at finite temperature, $\Gamma_0=0.1$, $T=0.2$.
Left, (a)-(c): 3-site model with nearest-neighbor tunneling, $J_1 =0.2$ and $J_2=0$.
(a) Population flux map. (b) Level diagram of the optimal configuration,
$\varepsilon_2= 0.21$, $\varepsilon_3= -0.034$, leading to $\eta=0.035$.
(c) Steady-state population of the optimal profile, \(\rho_{11}^{SS} = 0.40\), \(\rho_{22}^{SS} = 0.25\), \(\rho_{33}^{SS} = 0.35\).
Right, (d)-(e): 3-site model with next-nearest-neighbor tunneling, \(J_1 = 0.2\) and \(J_2 = 0.1\).
(d) Population flux map. 
(e) Level diagram of the optimal energy profile, obtained at \(\varepsilon_2 = -0.118\), \(\varepsilon_3 = -0.106\), and leading to
\(\eta = 0.033\). (f) Steady-state population of the optimal profile, \(\rho_{11}^{SS} = 0.36\), \(\rho_{22}^{SS} = 0.31\), \(\rho_{33}^{SS} = 0.33\).
Leak parameter is set at \(\gamma_l = 0.1\).
In both (a) and (d), the optimized structure from the OGA algorithm is presented by a black x.}
    \label{fig:figure4}
\end{figure*}


\section{Optimal Energy Landscapes in three-site systems} 
\label{sec:3siteoptim}

We begin by testing the optimization procedure on a minimal three-site model. Detailed analysis of such models were carried out in, e.g., Refs. \cite{Cao09,Liu2019}, with a particular focus on multi-path destructive interference processes.
In this setup, population exits from site 3 at a fixed rate constant
$\gamma_l$ and is injected back to site 1 at the same rate constant, establishing a steady state.
Without loss of generality, we set the first site energy, $\varepsilon_1$, to zero. We also fix the tunneling elements.
The remaining site energies, $\varepsilon_2$ and $\varepsilon_3$, serve as the optimization variables, with the objective of maximizing the population flux defined in Eq.~\eqref{eta}.
Since only two parameters are varied in the optimization process, we can visualize the flux landscape as a contour map. We can also easily compare the results of the optimization algorithm with the true global and local maxima obtained via extensive brute-force simulations.


\subsection{Coherent model}

We begin with the coherent case, obtained by setting \(\Gamma_n = 0\) in Eq. \eqref{lindpd}. Fig. \ref{fig:figure2} shows simulation results for the three-level system under two scenarios: a model with nearest-neighbor interactions only (left), and a model with an additional tunneling element between the first and third levels (right).
In both scenarios, we present a complete  flux map as a function of the two energy parameters, $\varepsilon_2$ and $\varepsilon_3$ while keeping $\varepsilon_1=0$. The point marked with an  ``x" indicates the scenario obtained via the OGA optimization protocol; it clearly successfully identified the optimal energy landscape for the system.

From Fig. \ref{fig:figure2}, we draw the following conclusions:
(i) When only nearest-neighbor coupling is allowed, i.e. $J_1\neq0$ and $J_2=0$ (left panels), the optimal energy landscape is nearly flat with $\varepsilon_{2,3}\approx \varepsilon_1$; see Fig. \ref{fig:figure2}(a)-(b).
Notably, a broad range of values for the intermediate level $\varepsilon_2$ supports close to optimal transfer fluxes.
In contrast, when the next-nearest-neighbor coupling is introduced, i.e. $J_1\neq0$, $J_2\neq0$, a V-shaped energy profile leads to enhanced transport and is optimal; see Fig. \ref{fig:figure2}(d)-(e).  
In this structure, the middle site (site 2) is essentially moved ``out of the way'', since the transfer can be facilitated by a direct tunneling transition from site 1 to site 3. 
Another notable observation is that optimal configurations show populations close to $1/3$. We recall that optimal flux corresponds to high population at the exit site (3), which in this system corresponds to a value approaching 1/3.

For the three-level model, one can solve the coherent scenario analytically in the steady-state limit. To simplify the expression, we consider the case $\varepsilon_1=\varepsilon_3=0$ while varying only $\varepsilon_2$ (in the simulations, $\varepsilon_3$ was also optimized). For nearest-neighbor tunneling, this yields
\bea
\eta(J_2=0,\Gamma=0) = \frac{4\gamma_l^2 J_1^4}{12 \gamma_lJ_1^4 + \gamma_l^3(\varepsilon_2^2+2J_1^2)}.
\label{eq:analgad0J20}
\eea
This simple result captures rich behavior: assuming that $\gamma_l \gg J_1$, that is, the leakage rate is not the slower, rate-determining step, we can distinguish between the deep tunneling limit, $\varepsilon_2\gg J_1$, and the case $\varepsilon_2\ll J_1$, 
which would evolve to band motion -- ballistic dynamics in long chains,
\bea
\eta(J_2=0,\Gamma=0; \gamma_l \gg J_1)=
\begin{cases}
\frac{4J_1^4}{\gamma_l\varepsilon_2^2} & {\rm for\,\,  }   \varepsilon_2\gg J_1\\
\frac{2J_1^2}{\gamma_l} &  {\rm for\,\,  }  \varepsilon_2\ll J_1.
\end{cases}
\eea
In the opposite limit of a small leakage rate, $\gamma_l\ll \varepsilon_2$ and $\gamma_l \ll J_1$, we find $\eta=  \gamma_l/3$, reflecting that the system reaches an equal-population state, and that
the flux is dictated by the leakage rate, which is the slowest, thus rate-determining, step. However, in our simulations, the parameters $\gamma_l$ and $J_1$ are of comparable magnitude. As such, neither of the limiting cases discussed above fully captures the observed behavior.

We now analyze analytically the role of next-nearest-neighbor tunneling, $J_2\neq 0$. 
The introduction of this tunneling element opens the door to interference effects. As before, we simplify the analysis by setting $\varepsilon_1=\varepsilon_3=0$, and neglecting environmental effects. In this case, the population flux is
\bea
&&
\eta(\Gamma=0)=
\nonumber\\
&&\frac{4\gamma_l^2 \left[ J_1^2-J_2(\varepsilon_2+J_2) \right]^2}{12 \gamma_l \left[ J_1^2-J_2(\varepsilon_2 +J_2) \right]^2 + \gamma_l^3\left[ \varepsilon_2^2+2\varepsilon_2J_2+ 2(J_1^2+J_2^2)\right]}.
\nonumber\\
\label{eq:analgad0}
\eea
When $J_2=0$, the flux is an even function of $\varepsilon_2$; the energy of the intermediate level can therefore be either positive or negative, see Eq. (\ref{eq:analgad0J20}). 
However, once $J_2\neq0$, this symmetry is broken. This is because the presence of $J_2$ introduces interference effects due to competing tunneling paths, and as such, the sign of $\varepsilon_2$ is influential. 
This effect is most clearly seen in the numerator of Eq. \eqref{eq:analgad0}, where the flux vanishes when
$J_1^2=J_2(\varepsilon_2+J_2)$. For example, choosing $J_1=0.2$ and $J_2=0.1$, the flux drops to zero at $\varepsilon_2=0.3$.
Conversely, negative values for $\varepsilon_2$ are expected to enhance the flux, as confirmed in Fig. 
\ref{fig:figure2}(d)-(e).
Once again we point out that our simulation parameters are such that both tunneling and leakage parameters are roughly of the same order, making transport mechanisms inherently complex and nontrivial to disentangle. 

The analysis of the coherent case in Fig. \ref{fig:figure2} provides three key guidelines for optimizing transport:
(i) In the absence of environmental effects there are many solutions (energy profiles)
 that are near optimal. These solutions cover a wide range of values for the intermediate level, $\varepsilon_2$. 
(ii) When multiple pathways are allowed due to next-nearest-neighbor tunneling, destructive interference effects appear when 
non-neighboring sites' energies are near resonance. To suppress these interference effects and achieve high flux, the intermediate level needs to be detuned from the entry and exit sites. That is, from Eq. (\ref{eq:analgad0}) we see that, under the assumption of  $J_2\ll J_1$, either setting $\varepsilon_2<0$ or $\varepsilon_2>J_1^2/J_2$ 
promotes transport.
(iii) The existence of multiple pathways, together with the openness of the system due to leakage, $\gamma_l\neq0$, 
breaks the symmetry of the energy profile with respect to $\varepsilon_2$, see Fig. \ref{fig:figure2}(d).

\subsection{OQS Model I}

We now incorporate environmental effects into the chain, in the form of local dephasing as described in Sec. \ref{sec:OQSI}. 
Figure \ref{fig:figure3} presents results for the three-level system, comparing the nearest-neighbor tunneling case [panels (a)–(c)] with the scenario that includes next-nearest-neighbor tunneling [panels (d)–(f)].
When only nearest-neighbor tunnelings are allowed (left), the flux depends on the interplay of coherent and incoherent effects. With the present parameters, this still favors a nearly flat energy profile, similar to the $\Gamma=0$ case of Fig. \ref{fig:figure2}(a)-(b).
When next-nearest-neighbor tunneling is included (right), interference effects take place, again similarly to the $\Gamma=0$ case, with the flux optimized near $\varepsilon_2\approx -0.3$.
Comparing Fig. \ref{fig:figure2} ($\Gamma=0$) with Fig. \ref{fig:figure3} ($\Gamma=0.1$),
we observe that in such short systems, transfer trends are similar, although in the latter case an asymmetry with respect to the energy $\varepsilon_3$ develops.

Next, we analyze the problem analytically by solving the Lindblad QME, Eq. (\ref{lindpd}) in the steady-state limit.
To simplify our analysis, we set $\varepsilon_1=\varepsilon_3=0$.
Excluding $J_2$, we derive the following result, which generalizes Eq. (\ref{eq:analgad0J20}),
\begin{widetext}
\bea
&&\eta(J_2=0)= 
\nonumber\\
&&\frac{2 \gamma_l J_1^2 \left[4 \Gamma^3+4 \Gamma^2 \gamma_l+\Gamma \left(\gamma_l^2+8 J_1^2\right)+2 \gamma_l J_1^2\right]}
{\Gamma^2 \left[12 \gamma_l \left(\varepsilon_2^2+4 J_1^2\right)+7 \gamma_l^3\right]+\Gamma \left[4 \gamma_l^2 \left(2 \varepsilon_2^2+5 J_1^2\right)+\gamma_l^4+48 J_1^4\right]+\gamma_l^3 \left(\varepsilon_2^2+2 J_1^2\right)+12 \Gamma^4 \gamma_l+8 \Gamma^3 \left(2 \gamma_l^2+3 J_1^2\right)+12 \gamma_l J_1^4}.
\nonumber\\
\label{eq:etaG}
\eea
\end{widetext}
When all parameters are of comparable magnitude, no single term dominates, making the full expression cumbersome and difficult to interpret. 
To gain some understanding, we expand Eq. (\ref{eq:etaG}) in orders of $\Gamma$. The first order correction to Eq. (\ref{eq:analgad0J20}) is given by
\bea
\eta (J_2=0)&=&
 \eta(J_2=0,\Gamma=0) 
\nonumber\\
&+&\frac{2 \gamma_l^2(\gamma_l^2\varepsilon_2^2J_1^2 - 8 \varepsilon_2^2J_1^4-12J_1^6)}
{(\gamma_l^2\varepsilon_2^2+2\gamma_l^2J_1^2+12J_1^4)^2}\Gamma + O(\Gamma^2)
\nonumber\\
\eea
Once again, we distinguish between the deep tunneling limit  ($\varepsilon_2\gg J_1$) and the near-resonant regime ($\varepsilon_2\ll J_1$), resulting in 
\bea
\eta(J_2=0)\approx
\begin{cases}
\frac{4J_1^4}{\gamma_l\varepsilon_2^2}  + 2\Gamma \frac{J_1^2}{\varepsilon_2^2} & {\rm for\,\,  }   \varepsilon_2\gg J_1\\
\frac{2J_1^2}{\gamma_l}  - \Gamma \frac{\gamma_l^2}{6J_1^2}  &  {\rm for\,\,  }  \varepsilon_2\ll J_1.
\end{cases}
\eea
As expected, environmental effects assist transport in the deep tunneling regime, but reduce the flux when transfer occurs near resonance.

In the opposite limit of large dephasing compared to the coherent energy parameters, $\Gamma\gg \varepsilon_2,J_{1},\gamma_l$ we get from Eq. (\ref{eq:etaG}) 
\bea
\eta(J_2=0)\approx \frac{J_1^2}{6\Gamma}.
\eea
The flux is inversely proportional to $\Gamma$, typical to the quantum Zeno effect or the (classical) large friction limit \cite{HanggiK}

We now take into account both environmental effects and next-nearest-neighbor tunneling.
For simplicity, we focus on the resonance case, $\varepsilon_{1,2,3}=0$.
We derive corrections to the flux up to second order in $J_2$ and first order in $\Gamma$:
\bea
\eta& \approx&  
\frac{4 \gamma_l (J_1^2 - J_2^2)^2}{12 (J_1^2 - J_2^2)^2 + 2 \gamma_l^2 (J_1^2 + J_2^2)}
- \Gamma \frac{\gamma_l^2J_1^2}{(\gamma_l^2+6J_1^2)^2} 
\nonumber\\
&+& 2 \Gamma \gamma_l^2 J_2^2
\frac{2\gamma_l^4 + 23 \gamma_l^2 J_1^2-42 J_1^4}{J_1^2(\gamma_l^2+6J_1^2)^3}
\label{eq:etaJ2Gad}
\eea
From the second term, we note that the interplay of dephasing and short-range ($J_1$) coherent tunneling is to suppress transport.
The last term is proportional to $J_2^2\Gamma$.
The sign of this term depends on the relative magnitude of $\gamma_l$ and $J_1$. As such, the interaction of environmental effects with the long-range $J_2$ coupling is either to enhance or suppress transport. Specifically, when $\gamma_l \ll J_1$ this last term becomes negative, reflecting destructive interference that reduces the flux.
However, it is important to remember that Eq. (\ref{eq:etaJ2Gad}) 
assumes a flat energy profile, and as such  does not capture the full complex picture that we present in 
Fig. \ref{fig:figure3} when $J_2\neq 0$.
This discussion is presented here to provide an appreciation of the range of transport regimes that appear in a system as short as three sites.

Finally, we point out that, as expected, Eqs. (\ref{eq:analgad0J20})-(\ref{eq:etaJ2Gad}) reveal that the flux depends on the leak rate, $\gamma_l$. This dependence is physical: the leak process represents an experimental extraction parameter that controls the rate at which population is collected from the exit site.
\subsection{OQS Model II} 

The local dephasing model, OQS Model I, can be regarded as an infinite temperature setup since the jump operators do not differentiate between energy excitation and relaxation processes in the energy basis. 
This is reflected in the steady-state solution, which reaches an equal population of $1/N$ across all sites for an $N$-site chain once $\gamma_l\to 0$, regardless of the energy profile.

To probe instead temperature-induced effects, we next study the three-site model at finite temperature, as described by OQS Model II, Eq. \eqref{lindT}. Results are presented in Fig. \ref{fig:figure4}, and we make the following observations:
(i) When $J_2=0$, the optimal energy profile is no longer flat. Instead, there is a preference for a structured energy profile, with $\varepsilon_2$ elevated above the other two sites and $\varepsilon_3 < \varepsilon_1$, see panels (a)-(b). At infinite temperature, or under OQS Model I, raising the intermediate level provides no advantage. However, at finite temperature under OQS Model II, this strategy enhances the flux. One way to rationalize this landscape is to point out that since the intermediate level is high in energy, its steady-state population is low (c), leaving more population available at the exit site 3.
(ii) When $J_2\neq0$, the strategy identified in Fig. \ref{fig:figure3} was to suppress destructive interference by lowering $\varepsilon_2$.
However, this approach is less effective at intermediate temperatures because it causes carrier accumulation at the middle site (2), rather than at the exit site.
As such, we find that the structure in Fig. \ref{fig:figure4} (d)-(e) 
balances interference effects with thermalization, yielding an optimal profile where $\varepsilon_2\sim\varepsilon_3<\varepsilon_1$.

\section{Optimal energy landscapes in long chains}
\label{sec:manysiteoptim}

Our aim is to build physical intuition and provide practical design principles for energy profiles that enhance carrier transport in quasi-one-dimensional chains. We focus on two key factors: (i) the interplay between coherent dynamics and environmental effects, and (ii) the influence of long-range versus short-range tunneling.

In Sec. \ref{sec:3siteoptim}, we demonstrated that the optimistic gradient ascent approach successfully identified the optimal energy profile for a three-site toy model. Building on this success, we now extend our analysis to longer chains. Instead of explicitly defining nearest- and next-nearest-neighbor tunnelings, we employ a power-law tunneling scheme, as described in Eq. \eqref{powerlaw}.
Two coupling regimes are studied: \(\alpha = 1\) and \(\alpha = 3\), corresponding to, respectively, ``long-range'' and ``short-range'' coupling.

In the following, we present results for nine- and ten-site systems. These cases were selected to investigate possible even-odd effects in the chain length. For completeness, Appendix \ref{AppendixB} provides parallel simulation results for intermediate chain lengths of five and six sites.
Given the complexity of the model, we did not attempt to solve it analytically. However, it is worth pointing out that mapping the quantum transport system to classical kinetic networks may allow derivations of closed-form expressions \cite{Cao09}.

In addition to considerations of coherent and incoherent effects, the choice of \(\gamma_l\) further controls this transport trend. However, to fairly compare results between different models, we keep this parameter fixed and instead focus on the roles of the inner-chain energy profile and the environmental model.

As a reminder, for an $N$-site chain, we optimize over the $N-1$ energy levels relative to site 1, which is fixed as a reference point to zero energy. Other parameters are kept constant. We set the maximum tunneling energy to $J_{\text{max}} = 0.2$ and the leak rate to $\gamma_l = 0.1$. 
In OQS Model I, the dephasing rate is set to $\Gamma = 0.1$, while in OQS Model II we use a temperature of $T = 0.2$ and set the dimensionless system-bath coupling to $\Gamma_0 = 0.1$.

\begin{figure*}[htbp]
\centering
\includegraphics[width=\linewidth]{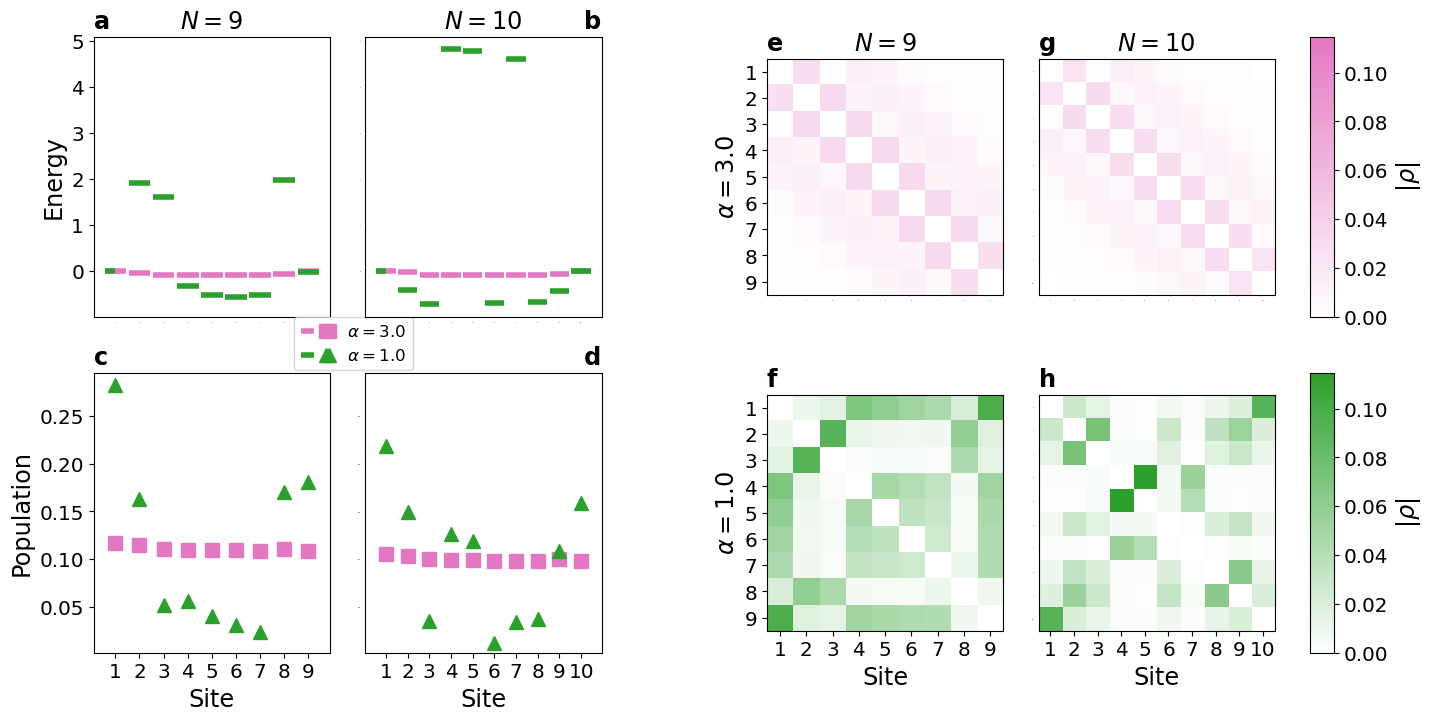}
\caption{Coherent Model: optimized energy landscape in (a) nine-site and (b) ten-site chains without environmental interactions, for short-range ($\alpha=3$) and long-range ($\alpha=1$) tunneling.
(a) $N=9$ sites profile with flux \(\eta_{\alpha=3} = 0.0109\) and \(\eta_{\alpha=1} = 0.0181\).
(b) $N=10$ sites profile with flux \(\eta_{\alpha=3} = 0.0098\) and  \(\eta_{\alpha=1} = 0.0159\).
Other parameters are \(J_{max} = 0.2\) and  $\gamma_l=0.1$.
(c)-(d) Steady-state populations for structures corresponding to (a)-(b).
(e)-(f) Absolute values of the steady-state density matrix elements  (diagonal removed) for the optimized structures in (a).
(g)-(h) Absolute values of the steady-state density matrix elements  (diagonal removed) for the optimized structures in (b).
Other parameters are \(J_{max} = 0.2\) and $\gamma_l=0.1$.}
    \label{fig:figure5}
\end{figure*}

\begin{figure}[htbp]
    \centering
\includegraphics[width=0.9\linewidth]{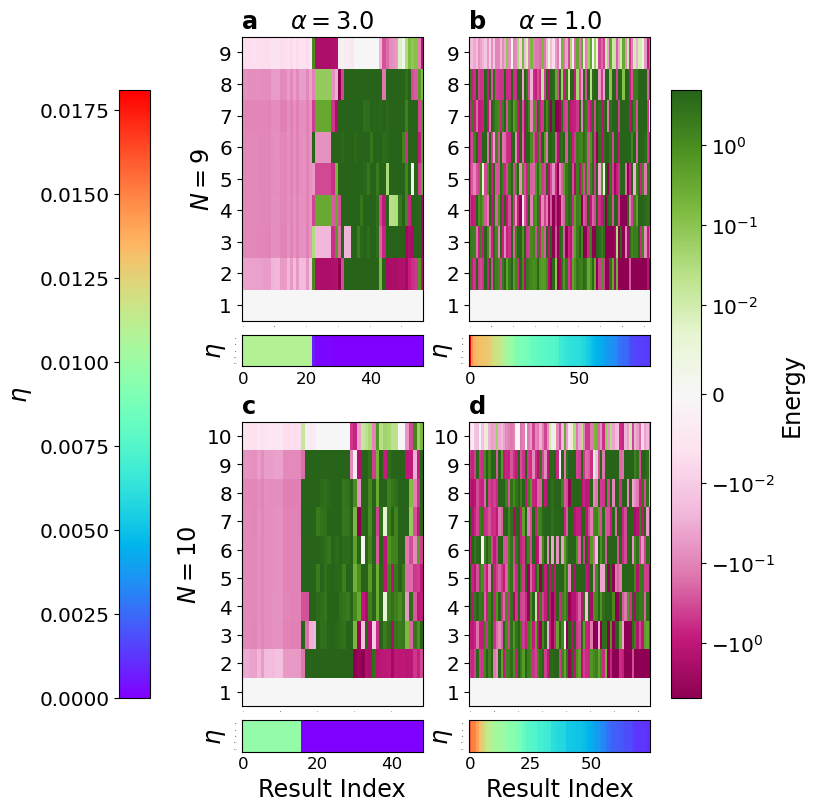}
\caption{Converged optimal energy profiles found when optimizing transport without environmental interactions, and their corresponding population flux. Each column in any panel is a single profile, and the leftmost columns are those plotted in Fig. \ref{fig:figure5}(a)-(b).
   Parameters are the same as in Fig. \ref{fig:figure5}.}  \label{fig:figure6}
\end{figure}

\subsection{Coherent Model}

Fig. \ref{fig:figure5}(a)-(b) show the energy profiles that optimize transport along the chain in the coherent limit.
For short-range tunneling ($\alpha=3$) the profiles remain nearly flat, reflecting an essentially uniform landscape. In contrast, the long-range tunneling case ($\alpha=1$) dramatically reshapes the energy landscape by ``pushing” intermediate energy levels away from resonance with both the entrance and exit sites, creating a more complex pattern. 

This reflects what we saw for the 3-site system in Eq. \eqref{eq:analgad0}, where a large, negative middle-site detuning led to greater transport under long-range tunneling (\(J_2 \neq 0\)). We intuit that a similar principle governs the transport behavior in longer chains, although in this case we see that the intermediate sites may be positively or negatively detuned.

The contrast between short-range and long-range tunneling is echoed in the population distributions shown in Fig. \ref{fig:figure5}(c)-(d).
To uncover the underlying transport mechanisms at play, we analyze in more detail the steady-state populations
and coherences under both coupling models. The steady-state populations for the short-range tunneling case ($\alpha=3$), shown in pink in Fig.~\ref{fig:figure5}(c)-(d), offer insight into why the corresponding energy profile is optimal. The population distribution exhibits a shallow gradient across the chain, indicative of quasi-ballistic transport, where the carriers flow with minimal scattering or localization.
In contrast, the steady-state populations for the long-range tunneling case ($\alpha=1$), shown in green in Fig.~\ref{fig:figure5}(c)-(d), reveal a strikingly different transport strategy. Here, population accumulates predominantly at the first and last sites, with suppressed occupation of the intermediate sites. This suggests that the optimal energy profile minimizes the occupation of intermediate sites by energetically detuning them, and helps avoid destructive interference effects, which are most pronounced when levels are near resonance. This proposed explanation on the mitigation of destructive interference effects could be examined in future work by decomposing the total flux into its pathways' components and analyzing the site-to-site contributions, following the approach of Ref. \cite{Liu2019}.
Though in the 10-site chain some intermediate levels surprisingly maintain high populations, generally the chain population exemplifies the competition between the classically-required population gradient for transport (see the explanation of ENAQT in \cite{zerah-harush2018, zerah-harush2020}), and our problem-specific requirement that the last site population be large so that the population flux is large.

The density matrix maps in Fig. \ref{fig:figure5}(e)-(h) corroborate the proposed transport mechanisms, with short-range tunneling ($\alpha=3$) facilitating transport through site-to-site coherent transport. In contrast, a direct tunneling between the entry and exit sites, or possibly assisted by some intermediate levels, is shown in the long range ($\alpha = 1$) case. 
The energetically-displaced levels exhibit strong coherence with each other, suggesting that they act as a ``bridge" that helps to preserve the carrier's coherence throughout the process.
However, we note that the steady-state populations in the coherent long-range model are highly sensitive to small variations in site energies. We suspect that the energy and population profiles depicted in Fig. \ref{fig:figure5} correspond to very sharp maxima in the population flux landscape, reflecting narrow ``sweet spots" where transport flux peaks; for more details, see Appendix \ref{sec:appAsharp}.



A natural question that arises is whether the optimized structures we obtain are constrained by our choice of initial optimizer conditions. In other words, are we truly finding the global maximum within the parameter space, or could there be even better solutions hidden beyond our current sampling?
In Figs. \ref{fig:figure6}(a)-(d) we present the successfully-optimized profiles; see Appendix \ref{AppendixA} for details on the optimization procedure. 
Although the optimization was repeated with many randomized initial ``guesses" for optimal profiles,
from Fig. \ref{fig:figure6}(a) and (c) 
we find that in the short-range case, many initial conditions reach the same optimal solution or closely similar solutions and, correspondingly, a similar flux. In contrast, for the long-range model, Fig.
\ref{fig:figure6}(b) and (d)
show that the parameter space apparently supports many local maxima. We identify several highly efficient solutions alongside a variety of sub-optimal ones. Notably, the best-performing solutions consistently share a key feature: the energies of the intermediate sites are strongly detuned from those of the entry and exit sites.

Figure \ref{fig:figure17} in Appendix \ref{AppendixB} demonstrates that five- and six-site chains display similar patterns in energy profiles, population distributions, and coherence dynamics across both short- and long-range tunneling regimes.
For short-range tunneling, the energy profile remains nearly uniform across the chain, hovering close to zero under our site 1 energy convention. In contrast, the long-range tunneling scenario favors setting the entrance and exit sites near resonance, while energetically detuning the intermediate sites. Interestingly, the most efficient solutions typically feature intermediate sites shifted to negative energies.
However, once again solutions under the long-range model are sensitive; small modifications in parameters can drastically alter the optimal energy landscape. This sensitivity gives rise to many locally-optimal energy profiles with intermediate sites detuned to positive energies, shown in Fig. \ref{fig:figure18}.

We summarize our findings on optimizing carrier transport in chains isolated from environmental effects. We refer to these as the ``Coherent model" (CM) design rules: for short-range tunneling, the optimal strategy favors an almost uniform energy landscape that supports quasi-ballistic transport. In contrast, when long-range tunneling plays a significant role, we propose that the optimal energy profile leverages these extended couplings while simultaneously suppressing destructive interference. This results in a non-uniform, corrugated energy landscape that energetically detunes intermediate sites to facilitate more direct transfer between the chain's endpoints. 

\begin{figure*}[htbp]
\includegraphics[width=\linewidth]{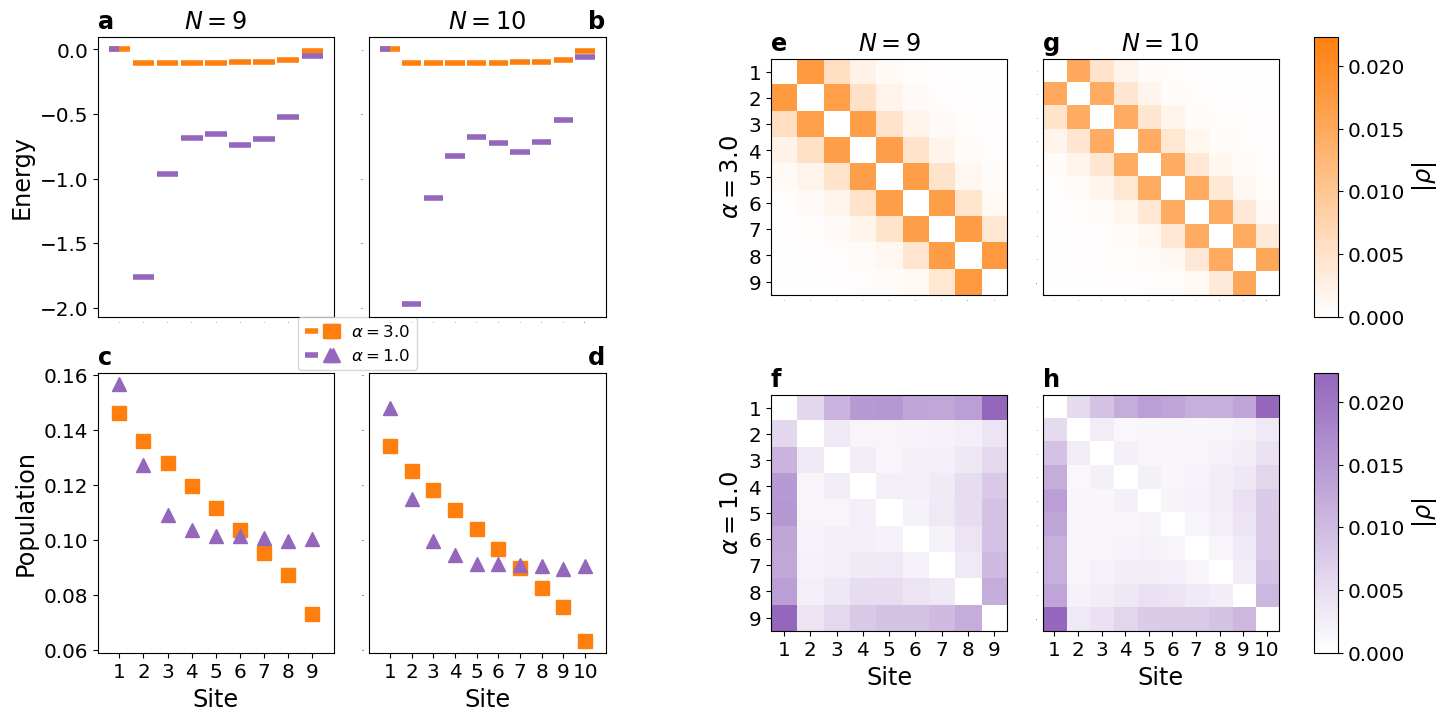}
    \caption{OQS Model I: optimized energy landscapes in (a) nine-site and (b) ten-site systems described by Eq. \eqref{lindpd} with $\Gamma=0.1$, for short-range ($\alpha = 3$) and long-range ($\alpha = 1$) tunneling.
(a) $N=9$ sites optimal profile with \(\eta_{\alpha=3} = 0.0073\) and \(\eta_{\alpha=1} = 0.0100\).
(b) $N=10$ sites optimal profile with \(\eta_{\alpha=3} = 0.0063\) and  \(\eta_{\alpha=1} = 0.0090\).   
(c)-(d) Steady-state populations corresponding to (a)-(b).
(e)-(f) Absolute values of the steady-state density matrix elements  (diagonal removed) for the optimized $N=9$ structures in (a). 
(g)-(h) Absolute values of the steady-state density matrix elements  (diagonal removed) for the optimized $N=10$ structures in (b). 
Other parameters are \(J_{max} = 0.2\) and $\gamma_l=0.1$.}
\label{fig:figure7}
\end{figure*}

\subsection{OQS Model I}

We extend our analysis to identify and classify chain configurations that maximize transport under varying conditions. 
Optimized structures for $N=9$ and $N=10$ chains under local dephasing are shown in Fig. \ref{fig:figure7} (a)–(b). Consistent with earlier observations, we find no significant even–odd effects: the optimal energy profiles for nine- and ten-site chains are strikingly similar. In all cases, the first and last sites exhibit comparable energies, mirroring the behavior seen in the minimal three-site model. 

Considering Fig. \ref{fig:figure7}(a)-(b), 
we first focus on the short-range tunneling scheme (orange). We observe that the largest energy gap occurs between sites 1 and 2. Beyond this, the site energies exhibit a slight non-monotonic increase, culminating in a more pronounced jump at the final site. The energy profile is remarkably similar to the case without the environment, Fig. \ref{fig:figure5}, although for $\Gamma\neq0$, the energy profile shows some spatial asymmetry around the center of the chain.  

For the long-range coupling scheme, the largest energy gap also occurs between sites 1 and 2, but it is approximately an order of magnitude larger than in the short-range case. Following another significant jump between sites 2 and 3, the energies of the intermediate sites increase gradually. Although the nine-site chains exhibit slightly higher population fluxes than the ten-site chains, their overall magnitudes remain comparable. Across both chain lengths, long-range coupling consistently enhances population flux, highlighting its significant advantage in facilitating carrier flow. 

To understand why these profiles are better than a uniform energy profile, it is helpful to return to the 3-site system results in Section \ref{sec:3siteoptim}. We see from equation \eqref{eq:etaJ2Gad} that
in the three-site system under OQS Model I,
a uniform energy profile leads to a decrease in transport with increasing $\Gamma$, when \(J_2 < J_1\). A similar effect likely applies to the longer 9- and 10-site chains; hence, it is not surprising that optimal energy profiles are not flat.

\begin{figure}[htbp]
    \centering \includegraphics[width=0.9\linewidth]{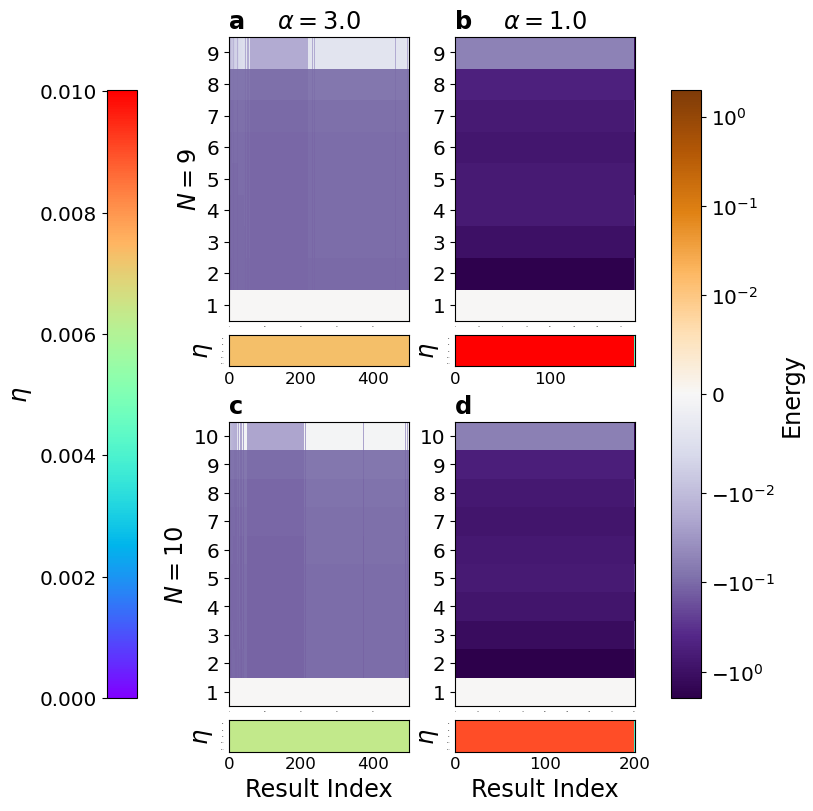}
\caption{
Converged optimal energy profiles found when optimizing transport in OQS Model I and their corresponding population flux. Each column in any panel is a single profile, and the leftmost columns are those plotted in Fig. \ref{fig:figure7}(a)-(b). Parameters are the same as in Fig. \ref{fig:figure7}.
}
\label{fig:figure8}
\end{figure}

\begin{figure}[htbp]
    \centering
\includegraphics[width=\linewidth]{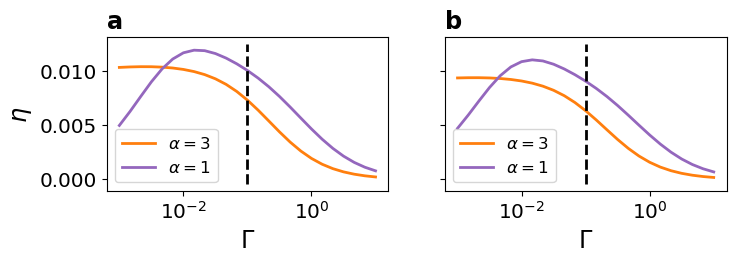}
\caption{Population flux \(\eta\) as a function of environmental noise \(\Gamma\) for 
(a) \(N = 9\)-site chain using the optimized energy profile for OQS Model I from Fig. \ref{fig:figure7}(a); 
(b) \(N = 10\)-site chain using the optimized energy profile for OQS I from Fig. \ref{fig:figure7}(b).
In the long range case, the corrugated energy landscape gives rise to an ENAQT effect. The dashed line is the fixed value of \(\Gamma\) used for the optimization.
}
\label{fig:figure9}
\end{figure}

\begin{figure*}[htbp]
    \centering
\includegraphics[width=0.95\linewidth]{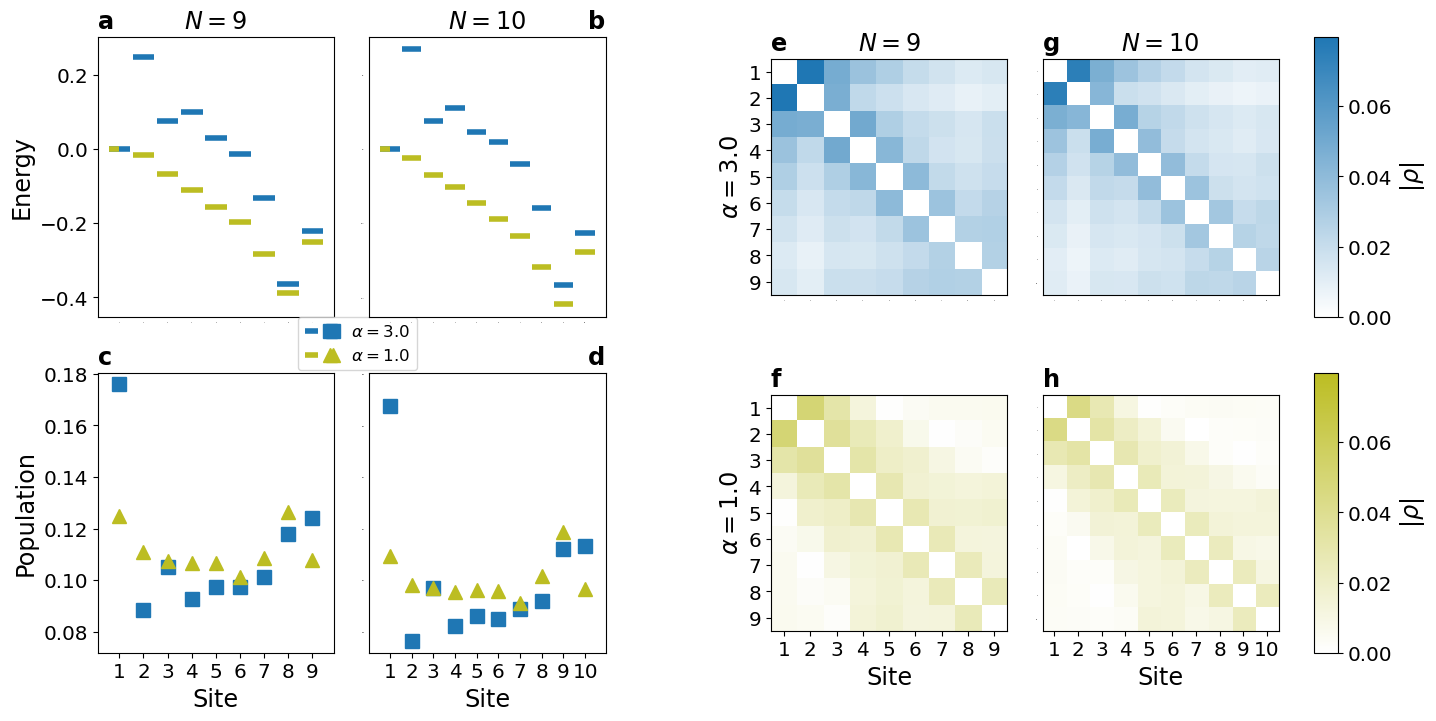}
\caption{OQS Model II: optimized energy landscapes in (a) nine-site  and (b) ten-site  chains coupled to a thermal bath with \(\Gamma_0 = 0.1\) and \(T = 0.2\), for short-range ($\alpha=3$) and long-range ($\alpha=1$) tunneling.
(a) $N=9$ sites profile with efficiencies \(\eta_{\alpha=3} = 0.0124\) and \(\eta_{\alpha=1} = 0.0108\).
(b) $N=10$ sites profile with efficiencies \(\eta_{\alpha=3} = 0.0113\) and  \(\eta_{\alpha=1} = 0.0097\).
(c)-(d) Steady state population corresponding to (a)-(b).
(e)-(f) Absolute values of the steady-state density matrix elements  (diagonal removed) for the $N=9$ optimized structures of (a).
(g)-(h) Absolute values of the steady-state density matrix elements  (diagonal removed) for the $N=10$ optimized structures of (b).
Other parameters are \(J_{max} = 0.2\) and $\gamma_l=0.1$.}
\label{fig:figure10}
\end{figure*}

\begin{figure}[htpb]
    \centering
\includegraphics[width=\linewidth]{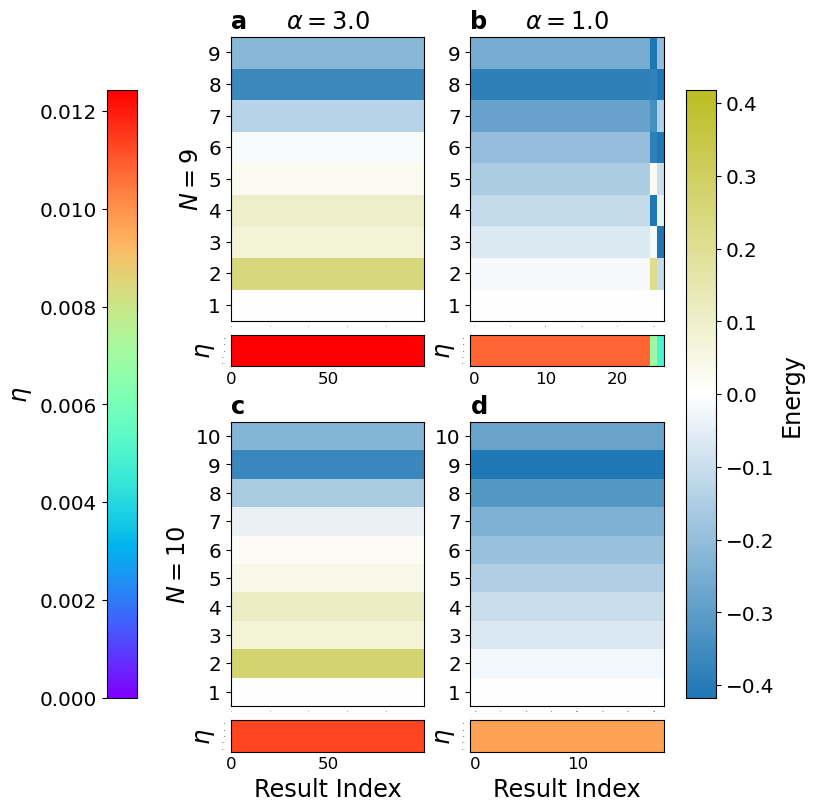}
\caption{Converged optimal energy profiles found when optimizing transport in OQS Model II and their corresponding population flux. Each column in any panel is a single profile, and the leftmost columns are those plotted in Fig. \ref{fig:figure10}(a)-(b). Parameters are the same as in Fig. \ref{fig:figure10}}
\label{fig:figure11}
\end{figure}


The optimal energy profiles in Fig. \ref{fig:figure7}(a)-(b) may appear complex to interpret. However, as with the coherent model, examining steady-state populations offers insight into why these structures facilitate optimal transport. Fig. \ref{fig:figure7}(c)-(d) present the steady-state populations at each site under both short- and long-range coupling schemes in the  nine- and  ten-site chains, respectively. 

Previous studies have shown that ENAQT in systems with nearest-neighbor tunneling arises from a competition between dephasing-driven population equalization and the requirement of a population gradient to drive flow from the excitation site to the sink, as classically described by Fick’s law \cite{zerah-harush2018, zerah-harush2020}. A pronounced population gradient signals predominantly classical transport behavior.
The optimal structures under short-range tunneling, shown in Fig. \ref{fig:figure7}(c)-(d) (orange squares) appear to conform to this classical picture: the 
population profile shows that there is nearly a constant population gradient in the chain, driving a diffusive flux, as described by Fick's law. 

To better understand the seemingly non-intuitive energy profiles observed in the long-range tunneling case, Fig. \ref{fig:figure7}(a)-(b), we turn to the corresponding steady-state populations in Fig. \ref{fig:figure7}(c)-(d). The population initially decreases, then saturates toward the end of the chain (purple triangles). 
Remarkably, 
the interplay between strong coherent effects enabled by large tunneling elements, and incoherent effects, enabled by comparable dephasing rates leads to a nontrivial energy structure that allows the system to establish an efficient transport, evidenced by the nearly constant population developing halfway through the chain.

Steady-state coherences in the site basis further shed light on the underlying transport mechanisms. Fig. \ref{fig:figure7}(e)-(h) displays the steady-state coherences for the optimal energy profiles shown in Fig. \ref{fig:figure7}(a)-(b). In the short-range case, significant coherences are observed only between nearest and next-nearest neighbors [Figs. \ref{fig:figure7}(e) and (g)] indicating that carrier transmission occurs predominantly via adjacent sites. 
In contrast, the long-range tunneling cases [Figs. \ref{fig:figure7}(f) and (h)] exhibit pronounced coherences between entry site 1 and several other remote sites, particularly the exit site $N$. It is interesting to note that site 2 shows only weak coherences with the rest of the chain, which aligns with its significantly lower energy relative to the other sites in the long-range energy profiles [Fig. \ref{fig:figure7}(a)-(b)]. 
Taken together with the steady-state populations, these findings suggest that, much like in the coherent case, long-range tunneling favors an optimal energy profile that promotes direct transport from the initial to the terminal site. By effectively bypassing the intermediate sites, this configuration enhances long-range transfer and boosts the overall flux.

Figure \ref{fig:figure8} shows the optimal energy profiles found, with their corresponding population flux. 
As before, each column is a single profile and the leftmost columns are those plotted in Figs. \ref{fig:figure7} (a) and (b). Although the optimization was repeated with many randomized initial ``guesses" for optimal profiles, Fig. \ref{fig:figure8} shows that each successful initial guess converged to the same profile under the OQS Model I. 


To complement our analysis and illustrate the related ENAQT behavior under our choice of parameters, Fig. \ref{fig:figure9} shows the population flux, \(\eta\), as a function of environmental noise. 
We use the structures previously identified as optimal for short- and long-range chains of nine and ten sites, and examine the flux while varying the dephasing rate, $\Gamma$. 
Notably, a turnover in flux, a hallmark of ENAQT, appears only in structures optimized with long-range tunneling, while the flux in purely short-range tunneling systems shows no such peak. This is expected given the localization dynamics that develop in corrugated structures, and subsequently incoherent effects serve to enhance transport. 
The flux peak in the long-range case does not align exactly with the $\Gamma$ value used to obtain the optimized energy profile, since the optimization was performed at a fixed $\Gamma$ (indicated by the dashed line). However, for the nine- and ten-site chains, the long-range peaks lie close to this value of $\Gamma$, confirming that the system operates within the ENAQT regime.

In Appendix \ref{AppendixB}, we repeat the optimization process of OQS Model I with shorter chains of five and six sites; see Fig. \ref{fig:figure19}. Our findings for these shorter chains are consistent with the nine- and ten-site chains: for models with short-range tunneling, we obtain nearly flat energy profiles and a constant gradient population. In contrast, long-range tunneling benefits from non-monotonic, corrugated energy landscapes, with population profiles exhibiting a clear transition from a decaying trend to an almost constant distribution along the chain.
 
We summarize our key findings on optimizing carrier transport in chains subject to local dephasing with the following OQS I design rule:
when local dephasing is strong and comparable in scale to the tunneling energies, the short-range tunneling model favors classical-like conduction with population decaying with length with a constant gradient. This is achieved through nearly flat energy landscapes with only a mild detuning from the entry and exit sites. In contrast, when long-range tunneling is allowed, optimal transport arises when the system successfully establishes quasi-ballistic conduction, as reflected by the steady-state population profiles. The observation of a relatively flat population profile at some sites away from the entrance ($n=4$ in Fig. \ref{fig:figure7}(c)-(d) and Fig. \ref{fig:figure19}(c)-(d)) under long-range tunneling is particularly striking. It demonstrates that efficient transport can emerge even in chains with highly non-uniform energy landscapes, provided that the energy structure is carefully balanced with the long-range tunneling amplitudes and environmental effects.  This transport regime is enabled by a highly nontrivial energy landscape that promotes efficient end-to-end transfer.


\subsection{OQS Model II}

We extend our investigation to transport optimization at finite temperatures using OQS Model II, as defined in Eq. \eqref{lindT}. The results are shown in Fig. \ref{fig:figure10}(a)–(b). Unlike the Coherent Model and OQS Model I, we observe consistent trends across both short- and long-range tunneling chains. In each case, the optimal energy profile adopts a ramp-like, steadily decreasing form. This configuration is physically intuitive: at moderate temperatures, it facilitates forward transfer from the entrance to the exit site while effectively suppressing backflow. 
For short-range tunneling, we also observe the emergence of high-energy levels above both the entrance and exit sites. Similarly to the three-site models shown in Figs. \ref{fig:figure4}(c) and (f), this arrangement reduces population buildup along the chain itself while enhancing accumulation at the exit site. 

Figure \ref{fig:figure10}(c)-(d) show the population distributions corresponding to the optimized energy profiles, while Fig. \ref{fig:figure10}(e)-(h) presents the off-diagonal elements of the density matrix.
Interestingly, after the first one or two sites, the population in the chain {\it increases} towards the exit site, indicating on the accumulation of carriers towards the end of the chain due to the decreasing ramp potential. 
Compared to optimal structures identified in OQS Model I, coherences in the finite-temperature model are more localized, extending only to about three neighboring sites for both short- and long-range tunneling cases. 
To highlight the robustness of these findings, Fig. \ref{fig:figure11} compiles a set of energy profiles achieved during the optimization process, all consistently exhibiting the characteristic ramp-like structure.

In Appendix \ref{AppendixB}, we repeat the optimization process of OQS Model II for chains with five and six sites, see Fig. \ref{fig:figure21}. We observe similar trends, with the energy levels following a ramp structure and the density matrix showing a relatively localized nature.

We summarize the key design principle for OQS Model II as follows: At moderate temperatures ($T\approx J_{max}$) and at weak system-environmental coupling $\Gamma_0T \ll1$, ramp-like, monotonically decreasing energy profiles are favored. These profiles support localized forward-moving carrier transport, and thus improve carrier collection at the exit site. Importantly, this design rule holds consistently across both short- and long-range tunneling models.
\begin{table*}[t!]
\begin{tabular}{|l|l|l|l|l|}
\hline
Model   & Figures & Energy landscape  & Population profile &  Coherences \\ 
\hline
 Coherent model, $\alpha=3$ & 
 \ref{fig:figure2}, \ref{fig:figure5}-\ref{fig:figure6}, \ref{fig:figure17}-\ref{fig:figure18} & nearly flat & nearly constant &  localized    \\
 \hline
  Coherent model, $\alpha=1$& \ref{fig:figure2}, \ref{fig:figure5}-\ref{fig:figure6}, \ref{fig:figure17}-\ref{fig:figure18} &corrugated &  bowl shaped & extended 
  \\ \hline
OQS Model I, $\alpha=3$ & \ref{fig:figure3},
\ref{fig:figure7}-\ref{fig:figure9}, 
\ref{fig:figure19}-\ref{fig:figure20} & nearly flat & decaying  approx. linearly &  localized  \\
\hline
OQS Model I, $\alpha=1$ &\ref{fig:figure3}, \ref{fig:figure7}-\ref{fig:figure9}, 
\ref{fig:figure19}-\ref{fig:figure20} & corrugated & 
constant for $n>4$ &  extended
\\ \hline
OQS Model II,  $\alpha=3$ & 
\ref{fig:figure4},\ref{fig:figure10}-\ref{fig:figure11}, \ref{fig:figure21}-\ref{fig:figure22}&  up then down ramp  & growing towards exit site&  localized 
\\ \hline
OQS Model II,  $\alpha=1$ & \ref{fig:figure4}, \ref{fig:figure10}-\ref{fig:figure11}, \ref{fig:figure21}-\ref{fig:figure22}& down ramp & approx. constant at  center&  localized 
\\ \hline
\end{tabular}
\label{tableM}
\caption{Summary of models, energy landscapes optimizing transport, and the associated proposed transport mechanism, deduced based on the behavior of coherences and the population profile. 
``Localized” coherences refer to cases where quantum coherences are strongest between nearest-neighbor sites and decay rapidly with distance. In contrast, ``extended” coherences describe situations where coherences with more distant sites can be comparable to or even stronger than those between immediate neighbors. 
}
\end{table*}

\begin{figure*}[htbp]
\includegraphics[width=\linewidth]{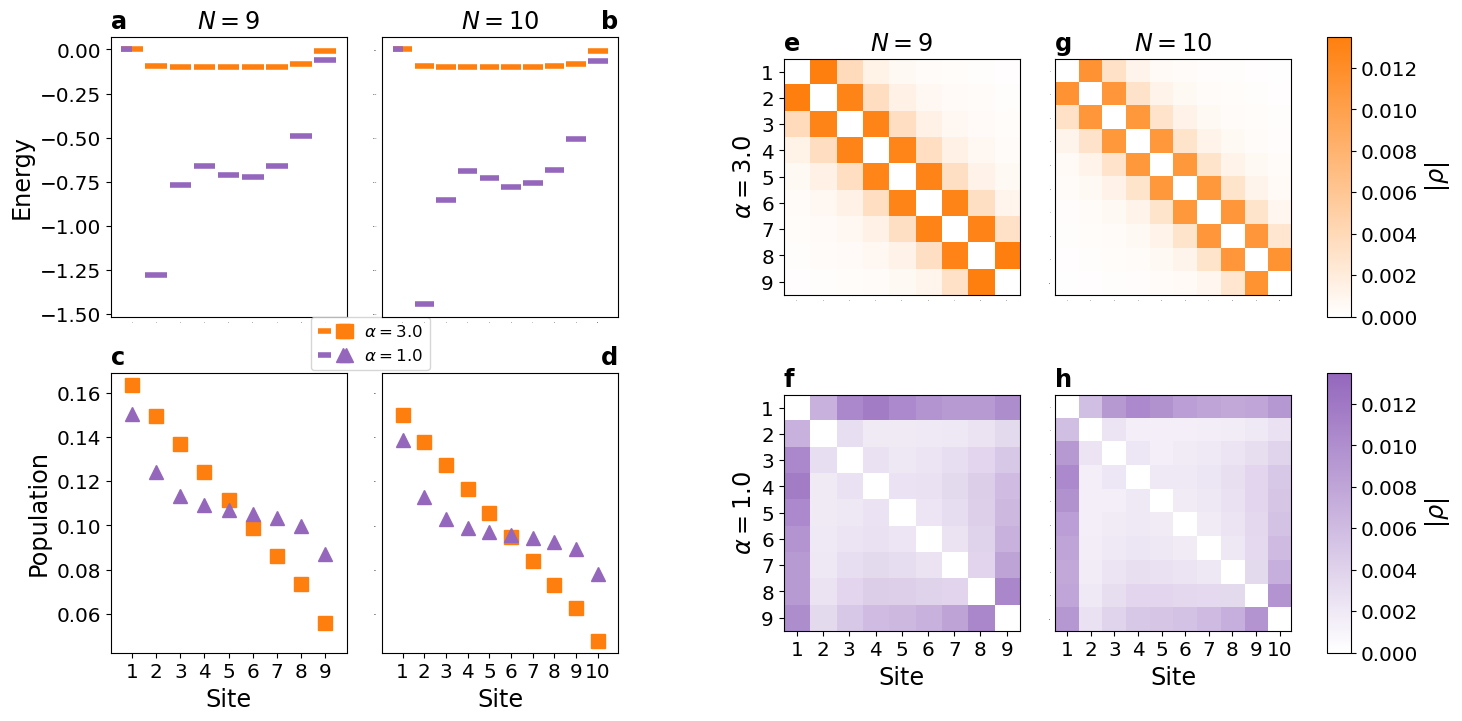}
    \caption{OQS Model I: Same as Fig. \ref{fig:figure7} but with a stronger dephasing constant, $\Gamma=0.2$.
%
Other parameters are \(J_{max} = 0.2\) and $\gamma_l=0.1$.}
\label{fig:figure12}
\end{figure*}

\begin{figure*}[htbp]
\includegraphics[width=\linewidth]{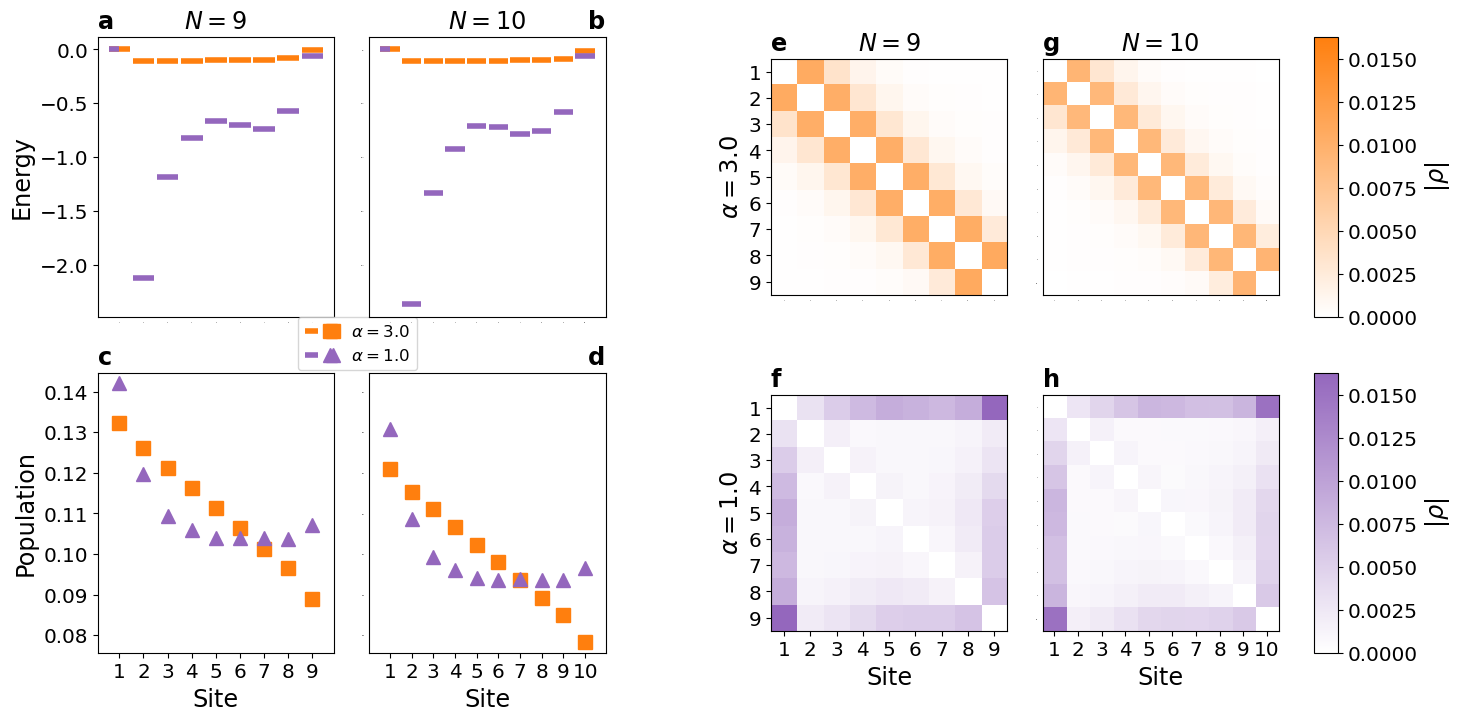}
    \caption{OQS Model I: Same as Fig. \ref{fig:figure7} but with a smaller leakage rate, $\gamma_l=0.05$.
%
Other parameters are \(J_{max} = 0.2\) and $\Gamma=0.1$.}
\label{fig:figure13}
\end{figure*} 

\subsection{Discussion}

We highlight that all optimization results were obtained under nonequilibrium steady state conditions and with a fixed value for the (physical) extraction rate $\gamma_l$. Our conclusions were drawn using a value for $\gamma_l$ comparable to other energy parameters. Having a very small $\gamma_l$ would turn it into the rate-determining step, while a very large $\gamma_l$ would effectively lead to a large level broadening at the exit site, shadowing variation in energy profile.


For OQS Model I, our analysis thus far has been performed using a fixed value of the environmental dephasing rate constant, $\Gamma$. It is important, however, to examine other values of $\Gamma$ to confirm the robustness of our observations. In Fig. \ref{fig:figure12}, we present additional results for OQS Model I, analogous to those in Fig. \ref{fig:figure7}, but obtained under stronger environmental noise. Overall, we find that the qualitative trends in the optimized energy profiles are preserved for both short-range and long-range tunneling systems. Interestingly, in the long-range tunneling case, the population profile at higher $\Gamma$
values shows a slight deviation from the flat behavior of  Fig. \ref{fig:figure7}, developing a small gradient across the bulk of the chain. This behavior aligns with the expectation that, as environmental effects become stronger relative to the tunneling amplitudes, long-range couplings, and their facilitation of coherent long-range hops, play a reduced role, giving way to more incoherent dynamics.
We also confirmed convergence of the structure reported
in Fig. \ref{fig:figure12} for many initial values, as in Fig. \ref{fig:figure8}, and that the long-range system supports an ENAQT regime at $\alpha=1$, similarly to Fig. \ref{fig:figure9}.

This example reinforces the rough classification of three transport regimes for the system, as discussed in Sec. \ref{sec:transport}: coherent, incoherent and intermediate. Our choice of parameters throughout this project has been to achieve the intermediate regime, thus allowing nontrivial competition and optimization. The comparison of Fig. \ref{fig:figure12} with Fig. \ref{fig:figure7} clearly shows that increasing the dephasing effects relative to coherent effects takes us further from the intermediate regime to the incoherent limit.

We further examine the impact of changing $\gamma_l$ on the OQS Model I results. In Fig.
\ref{fig:figure13}, we use $\gamma_l=0.05$ 
compared to $\gamma_l=0.1$ in Fig.
\ref{fig:figure7}.
We find that modification of the leak rate does not fundamentally change the results but that a smaller $\gamma_l$ leads to the accumulation of population near the exit site, which is expected.
In contrast, we found that an increase in $\gamma_l$ (not shown) led to the depletion of the population at the exit site, compared to the bulk.

Overall,
Figs. \ref{fig:figure7},
\ref{fig:figure12} and
\ref{fig:figure13}
demonstrate the robustness of our observations when environmental noise or leak rate are varied, yet kept within the range of other parameters. Particularly we note that the optimization results are more robust when interactions are short-range. 


\section{Conclusions}
\label{sec:summary}

Using modern optimization algorithms, we have identified distinct classes of energy landscapes that maximize population flux along quasi-one-dimensional chains. Hand-in-hand with the optimization, the analysis of the resulting steady state has allowed us to guess at underlying transport mechanisms. 
Toward the objective of optimized transport, we specifically studied (i) the role of the environment and (ii) the impact of short-range vs. long-range tunneling on chain design. The interplay of these incoherent and coherent effects leads to complex transport behavior. Optimized transport in the models required these factors to cooperate. 
Importantly, in the presence of long-range tunneling couplings, the optimization approach discovered energetic profiles that would be very difficult to guess analytically.

We modeled the chain’s interaction with its environment using the Lindblad QME. In OQS Model I, we incorporated local Lindblad dephasing, which, when viewed in the energy eigenbasis, corresponds to coupling with an infinite-temperature bath. In contrast, OQS Model II represents a finite-temperature environment that drives both excitation and relaxation processes between energy eigenstates, while satisfying the detailed balance condition.

We have identified the following design rules for optimizing transfer, summarized in Table I.

(i) Coherent model: for short-range tunneling, the optimal configuration features an almost uniform energy landscape, enabling a constant population profile and thus near-resonant transport. 
When long-range tunneling contributes, the optimal energy profile becomes highly non-uniform and corrugated, with intermediate sites energetically detuned. This facilitates a more direct transfer between the chain's endpoints, as evidenced by strong steady-state coherences linking those sites. Long-range {\it coherent} tunneling systems are highly sensitive to variations in parameters near optimal energy landscapes.

(ii) OQS Model I: when local dephasing is comparable in magnitude to the tunneling coupling energies, models with short-range tunneling favor nearly flat energy landscapes, with only slight detuning from the entry and exit sites to facilitate conduction. The population profile in this case suggests diffusive transport.
In contrast, under long-range tunneling, a corrugated energy landscape allows the system to cross over to quasi-ballistic conduction away from the entrance site by suppressing interferences and facilitating efficient end-to-end transfer. The development of this motion under highly corrugated structures and with dephasing effects manifests that the optimization approach can discover patterns that would otherwise be impossible to determine analytically. 

(iii) OQS Model II: at moderate temperatures ($T\approx J_{max}$), ramp-like energy profiles with a total gap of the order of $J_{max}$ are preferred, 
as they support directional transport and improve the collection of carriers at the exit site. In these systems, regardless of the range of the tunneling couplings, transport proceeds in an almost localized manner with small coherences, and with population concentrating towards the bottom (end) of the ramp. 

We emphasize that the transport mechanisms discussed in this work are inferred only from the observed patterns of coherences and population profiles. Because the energy landscape was optimized separately for structures of different lengths, potentially leading to distinct profiles for different sizes, we cannot systematically examine how the flux scales with system size for identical Hamiltonians. Thus, a quantitative analysis of transport mechanisms is beyond the scope of this study. Our objective, instead, has been to identify families of systems that enhance flux under different conditions, rather than to classify transport mechanisms in detail.

The optimization process posed challenges at times, especially for the Coherent Model and OQS Model II. In the coherent model, particularly with long-range tunneling, the steady-state populations were highly sensitive to small changes in energy levels, resulting in complex, rugged energy landscapes with numerous local maxima that complicated optimization. For OQS Model II, the need to re-diagonalize the system Hamiltonian at each energy iteration further increased computational demands, making the optimization considerably slower compared to OQS Model I under the same parameters.

The concept of ENAQT was originally introduced to explain the efficiency of light-harvesting systems \cite{Alan08,Plenio08,Plenio09, Plenio10}. It was later generalized to broader scenarios of transport in low-dimensional chains \cite{maier2019environment,Blach25}. We believe that the main impact of our work lies in providing a deeper understanding of transport in such engineered setups, with potential relevance for applications such as photovoltaics.
Long-range steady-state coherences, such as those found under OQS Model I in Fig. \ref{fig:figure7}(f)-(h) have been shown to lead to robust entanglement \cite{Streltsov2017Coherence, Dutta2020CoherenceSteadyStates, Kim2022coherence}. As such, the energy profiles we obtain under long-range tunneling [purple in Fig. \ref{fig:figure7}(a)-(b)] 
may find applications in quantum information processing and quantum metrology \cite{Dutta2020CoherenceSteadyStates, Schindler2013QuantumMapsIons, Katsumi2025HybridDiamond}.

Future work could extend this study of carrier transfer to multi-carrier scenarios, such as transport in spin systems modeled by Heisenberg-type chains with long-range exchange interactions. Another promising avenue is the exploration of excitation transfer in molecular aggregates or disordered materials coupled to optical cavities \cite{Polariton,Pupillo15,Cao22P,David23,ScholesWu}, where cavity interactions mediate long-range transfer. Joint optimization of both the molecular structure and its cavity coupling may uncover new design principles for efficient carrier transport in quantum networks.

In sum, this work establishes a framework for designing quantum systems that leverage the synergy of coherent and incoherent processes to optimize transport performance. Our results provide guiding principles for engineering energy profiles to achieve efficient transport, with potential applications in organic photovoltaics, nanoelectronics, and quantum networks. 

\begin{acknowledgments}
We thank Roeland Wiersema for useful discussions. The work of M.L. and M.P. is supported by the NSERC Canada Graduate Scholarship-Doctoral. E.F. was funded by the 
Natalia Krasnopolskaia Summer Undergraduate Research Fellowships of the Department of Physics, University of Toronto.
D.S. acknowledges support from an NSERC Discovery Grant and an NSERC Alliance International Catalyst Grant. 
Resources used in preparing this research were provided, in part, by the Province of Ontario, the Government of Canada through CIFAR, and companies sponsoring the Vector Institute \url{www.vectorinstitute.ai/#partners}. 
\end{acknowledgments}

\vspace{4mm}
\noindent {\it Data Availability.}
The data that support the findings of this article are openly available \cite{Data}. The underlying code will be made available early in 2026.

\appendix


\section{Details of the optimization process}
\label{AppendixA}

\begin{figure}[h]
    \centering
    \includegraphics[width=\linewidth]{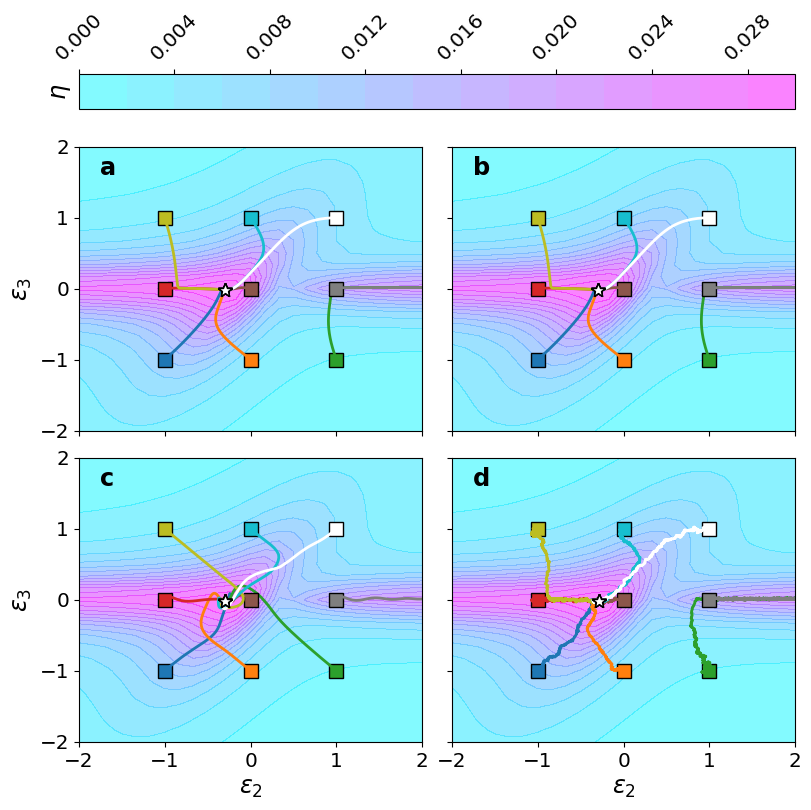}
    \caption{Examples of optimization trajectories in a three-site system under OQS Model I. Squares mark the initial energies supplied to the optimizer, while stars indicate where the optimizers reached their stopping condition. Default Optax parameters were used for each algorithm, except for the learning rate. The optimizers are: (a) optimistic gradient ascent, learning rate 0.5; (b) stochastic gradient descent, learning rate 0.2; (c) AdaMax, learning rate 0.05; (d) noisy stochastic gradient descent,  learning rate 0.2. Model parameters are \(J_1 = 0.2\), \(J_2 = 0.1\), \(\Gamma = 0.1\), \(\gamma_l = 0.1\). }
    \label{fig:figure14}
\end{figure}

Although OGA and AdaMax were used to perform the optimizations, other gradient-based optimization algorithms were also considered.  Some demonstrations of this with three sites are shown in Fig. \ref{fig:figure14}, with their default Optax parameters --- each found the true optimal values \((\varepsilon_2 = -0.292, \varepsilon_3 = -0.017)\) for most initial energies.  Ultimately, OGA and AdaMax were chosen because the optimizations terminated successfully in a relatively small number of steps. In Fig. \ref{fig:figure14}(a), the grey and green trajectories failed to find a maximum within the allowed range of energies, but the other trajectories took between 361 steps (brown) and 862 steps (red). The algorithms in Fig. \ref{fig:figure14} were tested with their default Optax hyperparameters except the learning rates, which are given in the caption. In theory, the default hyperparameters could be tuned as well, potentially leading to better convergence trends.

Many gradient descent algorithms used in machine learning/neural networks have a ``momentum'' term incorporated into their update step to help them cross ``barren plateaus'' where gradients are small. However, for many of our systems, our target maxima can be long ridges, which led most algorithms with momentum to find solutions farther along the ridge than they needed to. However, we found that AdaMax was better at finding maxima in complex flux landscapes where OGA failed due to its momentum term; comparing Fig. \ref{fig:figure14}(c) with (a), it is clear that AdaMax manages to explore the landscape more broadly.

Stochastic Gradient Descent (SGD) and Noisy SGD algorithms were also tested with three sites, the results of which are shown in Figs. \ref{fig:figure14}(b) and \ref{fig:figure14}(d), respectively. Both found the same optima as OGA and AdaMax, but with more steps: SGD took between 890 and 17,700 steps, and with the default noise levels, Noisy SGD took between 19,000 and 28,000 steps.

We emphasize that our work does not employ, nor does it develop, machine learning models. Instead, we use gradient-based optimization methods, specifically, gradient ascent, whose efficiency is enhanced through automatic differentiation (AD).
Although AD was developed alongside modern machine learning, it is a general computational technique for exactly and efficiently propagating derivatives through general programs via the chain rule. It enables rapid and accurate evaluation of gradients, therefore automating and accelerating numerical optimization. However, it does not involve statistical learning or model training.

Differentiation through eigendecomposition in non-hermitian matrices is not implemented in JAX. Therefore, for OQS Model I, it was more practical to solve for the steady-state density matrix as described in Section \ref{sec:SSsolve}, rather than by diagonalizing the matrix $M$ and taking the eigenvector corresponding to the eigenvalue equal to zero, as described in \cite{dey2022}. For OQS Model II, where the Lindblad QME is expressed in the Hamiltonian eigenbasis, the eigendecomposition function in the Python Library FMMAX \cite{schubert2023fourier} was used due to its compatibility with JAX's automatic differentiation. The steady-state density matrix was still solved for in the manner described in Section \ref{sec:SSsolve} to slightly accelerate calculations.

JAX's functions are multithreaded by default, making them incompatible with other Python parallelization libraries such as Python's native multiprocessing library, or Joblib \cite{joblib}. We had good success using JAX's vectorization function, vmap, to vectorize our optimization loop over all initial sets of site energies instead.

Model-specific details of the optimization are presented below. All initial energy profiles supplied to the optimizers (OGA or AdaMax) were uniformly sampled from a hypergrid of energies \(\{\varepsilon_i\} \in [-1, 1]\), for \(i \in [2, N]\). For \(N = 3\), 40 points were generated per energy dimension. For larger systems, 4 points were generated per energy dimension.

\subsection{Coherent Model} \label{sec:AppACM}

In the coherent model, the AdaMax algorithm was used with a learning rate of 0.1 to optimize all chain lengths. For each, a maximum step count of 100,000 was imposed, and 100 trials were performed. In what follows, we use the term ``converged" to refer to the trials that reached a local or global maximum in fewer iterations than the step count limit.

For 3-site systems, the converged energy profiles are shown in Fig. \ref{fig:figure2}(a) and (d). In the nearest-neighbor tunneling case ($J_1 = 0.2$, $J_2 = 0$) all 100 initial energy profiles successfully converged to the same maximum in fewer than 1,700 steps, shown in Fig. \ref{fig:figure2}(b). In the next-nearest-neighbor tunneling case ($J_1 = 0.2$, $J_2 = 0.1$) 67 out of 100 initial energy profiles successfully converged to the same maximum in fewer than 1,300 steps, shown in Fig. \ref{fig:figure2}(e).

For the 9-site systems, the converged energy profiles are shown in Fig. \ref{fig:figure6}(a) and (b). The short-range tunneling ($\alpha = 3$) model terminated successfully in 57 out of 100 trials, all in under 96,000 steps. The long-range tunneling ($\alpha = 1$) model terminated successfully in 83 out of 100 trials, all in under 96,000 steps.

For the 10-site systems, the converged energy profiles are shown in Fig. \ref{fig:figure6}(c) and (d). The short-range tunneling ($\alpha = 3$) model terminated successfully in 49 out of 100 trials, all in under 90,000 steps. The long-range tunneling ($\alpha = 1$) model terminated successfully in 75 out of 100 trials, all in under 99,000 steps.

For the 5-site systems, the converged energy profiles are shown in Fig. \ref{fig:figure18}(a) and (b). The short-range tunneling ($\alpha = 3$) model terminated successfully in 99 out of 100 trials, all in under 13,000 steps. The long-range tunneling ($\alpha = 1$) model terminated successfully in all 100 trials, all in under 2,000 steps.

For the 6-site systems, the converged energy profiles are shown in Fig. \ref{fig:figure18}(c) and (d). The short-range tunneling ($\alpha = 3$) model terminated successfully in 82 out of 100 trials, all in under 53,000 steps. The long-range tunneling ($\alpha = 1$) model terminated successfully in 98 out of 100 trials, all in under 93,000 steps.

\subsection{OQS Model I} \label{sec:AppAI}

For OQS Model I, the Optimistic Gradient Ascent algorithm was used with a learning rate of 0.5 to optimize all chain lengths. For each, a maximum step count of 100,000 was imposed. Being the first optimizations we attempted, 500 trials were performed for 3-, 9-, and 10-site systems, but given their repeated success, it was decided that 100 trials were sufficient to discover solutions of interest for all other systems.

For the 3-site systems, the converged energy profiles are shown in Fig. \ref{fig:figure3}(a) and (d). In the nearest-neighbor tunneling case ($J_1 = 0.2$, $J_2 = 0$) all 500 initial energy profiles successfully converged to the same maximum in fewer than 10,000 steps, shown in Fig. \ref{fig:figure3}(b). In the next-nearest-neighbor tunneling case ($J_1 = 0.2$, $J_2 = 0.1$) 423 out of 500 initial energy profiles successfully converged to the same maximum in fewer than 10,000 steps, shown in Fig. \ref{fig:figure3}(e).

For the 9-site systems, the converged energy profiles are shown in Fig. \ref{fig:figure8}(a) and (b). The short-range tunneling ($\alpha = 3$) model terminated successfully in 499 out of 500 trials, all in under 18,000 steps. The long-range tunneling ($\alpha = 1$) model terminated successfully in 190 out of 500 trials, each in under 38,000 steps.

For the 10-site systems, the converged energy profiles are shown in Fig. \ref{fig:figure8}(c) and (d). The short-range tunneling ($\alpha = 3$) model terminated successfully in 499 out of 500 trials, each in under 25,000 steps. The long-range tunneling ($\alpha = 1$) model terminated successfully in 201 out of 500 trials, each in under 52,000 steps.

For the 5-site systems, the converged energy profiles are shown in Fig. \ref{fig:figure20}(a) and (b). The short-range tunneling ($\alpha = 3$) model terminated successfully in all 100 trials, each in under 2,000 steps. The long-range tunneling ($\alpha = 1$) model terminated successfully in 69 out of 100 trials, each in under 5,000 steps.

For the 6-site systems, the converged energy profiles are shown in Fig. \ref{fig:figure20}(c) and (d). The short-range tunneling ($\alpha = 3$) model terminated successfully in all 100 trials, each in under 3,000 steps. The long-range tunneling ($\alpha = 1$) model terminated successfully in 62 out of 100 trials, each in under 13,000 steps.

\subsection{OQS Model II} \label{sec:AppAII}

For OQS Model II, OGA was used with 3 sites with a learning rate of 0.05. The AdaMax algorithm was used with a learning rate of 0.05 to optimize chains of five, six, nine, and ten sites. For each, 100 trials were performed, each with a maximum step count of 500,000; due to the small learning rate, more steps were allowed, although this ultimately proved unnecessary. AdaMax was used for the longer systems due to difficulties getting OGA to converge.

For three sites, the converged energy profiles are shown in Fig. \ref{fig:figure4}(a) and (d). In the nearest-neighbor tunneling case ($J_1 = 0.2$, $J_2 = 0$) all 100 initial energy profiles successfully converged to the same maximum in fewer than 7,200 steps, shown in Fig. \ref{fig:figure4}(b). In the next-nearest-neighbor tunneling case ($J_1 = 0.2$, $J_2 = 0.1$), 73 out of 100 trials successfully converged to the same maximum in fewer than 9,000 steps, shown in Fig. \ref{fig:figure4}(e).

For the 9-site systems, the converged energy profiles are shown in Fig. \ref{fig:figure11}(a) and (b). The short-range tunneling ($\alpha = 3$) model terminated successfully in all 100 trials, each in under 1,700 steps. The long-range tunneling ($\alpha = 1$) model terminated successfully in 27 out of 100 trials, each in under 5,000 steps.

For the 10-site systems, the converged energy profiles are shown in Fig. \ref{fig:figure11}(c) and (d). The short-range tunneling ($\alpha = 3$) model terminated successfully in all 100 trials, mostly in under 4,000 steps, except for one which took near 282,000 steps. The long-range tunneling ($\alpha = 1$) model terminated successfully in 19 out of 100 trials, each in under 2,100 steps.

For the 5-site systems, the converged energy profiles are shown in Fig. \ref{fig:figure22}(a) and (b). The short-range tunneling ($\alpha = 3$) model terminated successfully in all 100 trials, each in under 310 steps. The long-range tunneling ($\alpha = 1$) model terminated successfully in 53 out of 100 trials, each in under 170 steps.

For the 6-site systems, the converged energy profiles are shown in Fig. \ref{fig:figure22}(c) and (d). The short-range tunneling ($\alpha = 3$) model terminated successfully in all 100 trials, each in under 420 steps. The long-range tunneling ($\alpha = 1$) model terminated successfully in 31 out of 100 trials, each in under 1,700 steps.

\begin{figure}
    \centering
    \includegraphics[width=1\linewidth]{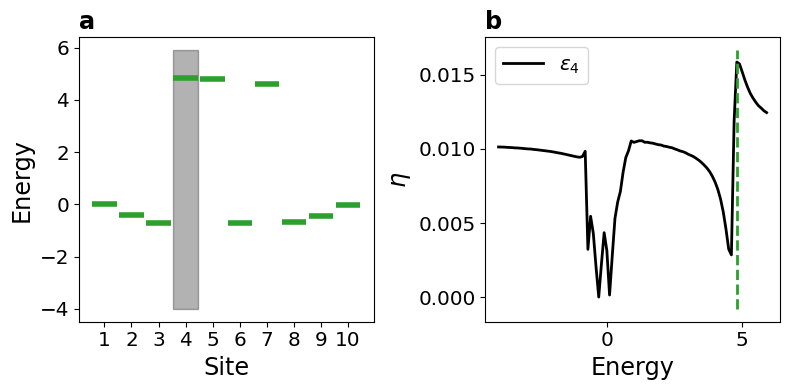}
    \caption{(a) Coherent Model: the optimized energy landscape shown in Fig. \ref{fig:figure5}(b) for long-range (\(\alpha=1\)) tunneling. The black bar shows the range over which \(\varepsilon_4\) was varied. (b) Population flux \(\eta\) as a function of \(\varepsilon_4\). The dashed line shows the value of \(\varepsilon_4\) found through the optimization procedure.
    Other parameters are \(J_{max} = 0.2\) and \(\gamma_l = 0.1\).}
    \label{fig:figure15}
\end{figure}

\begin{figure}
    \centering
    \includegraphics[width=1\linewidth]{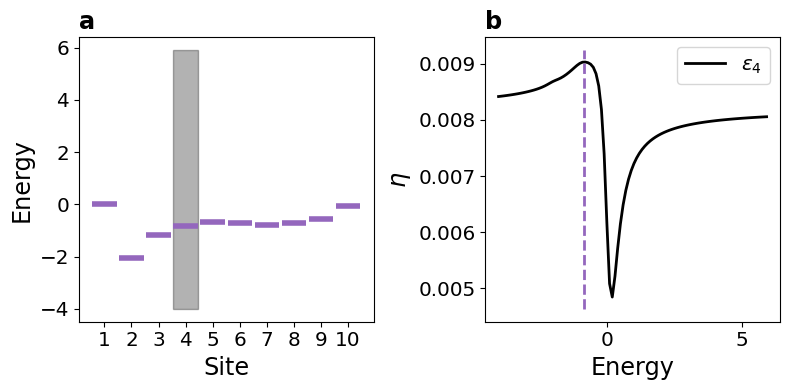}
    \caption{(a) OQS Model I: the optimized energy landscape shown in Fig. \ref{fig:figure7}(b) for long-range (\(\alpha = 1\)) tunneling. The black bar shows the range over which \(\varepsilon_4\) was varied. (b) Population flux \(\eta\) as a function of \(\varepsilon_4\). The dashed line shows the value of \(\varepsilon_4\) found through the optimization procedure. Other parameters are \(J_{max} = 0.2\), \(\Gamma = 0.1\), and \(\gamma_l = 0.1\).}
    \label{fig:figure16}
\end{figure}

Sharpness of Maxima \label{sec:appAsharp}
Optimizing the system energies under the Coherent Model (CM) of transport was difficult. These optimizations took more steps, and fewer initial energy configurations successfully converged to local maxima. To confirm our optimal profile was indeed a maximum, we study the effects of varying \(\varepsilon_4\) in the best 10-site, long-range tunneling system optimized under the CM while leaving the other energies fixed at their optimal values. This is shown in Figure \ref{fig:figure15}. The maximal population flux does indeed occur at the value of \(\varepsilon_4\) found by the Adamax algorithm, though this peak is very narrow, with a steep slope to its left. There are other local maxima near \(\varepsilon_4 \approx 0\), though they are also sharply peaked. In contrast, when we perform the same variation in \(\varepsilon_4\) with the best 10-site, long-range tunneling system optimized under OQS I, shown in Figure \ref{fig:figure16}, we see a much smoother maximum and no other local maxima in the range of energies considered.


\section{Optimal energy landscapes: $N=5$ and $N=6$-site chains}
\label{AppendixB}




\begin{figure*}[htpb]
\centering
\includegraphics[width=0.9\linewidth]{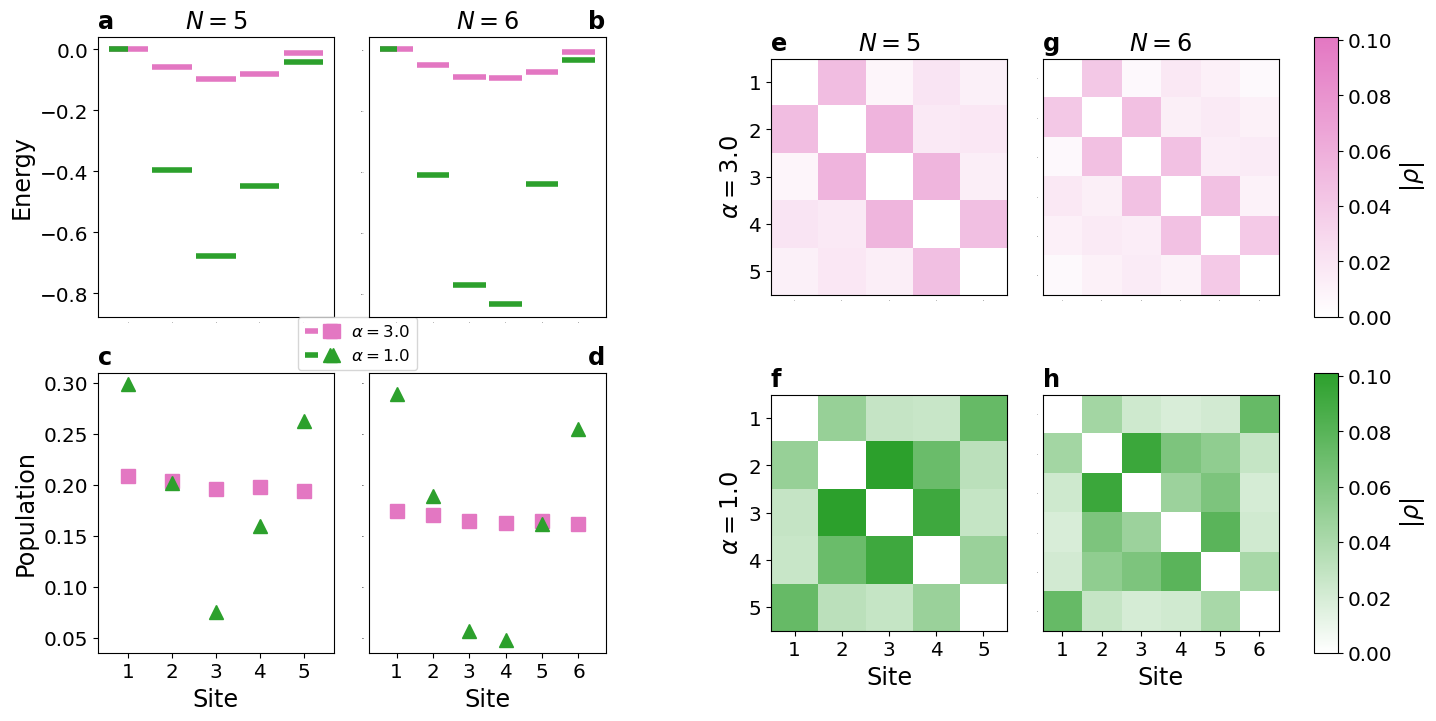}
    \caption{Coherent Model: optimized energy landscapes in (a) five-site and (b)  six-site chains without environmental interactions, for short-range ($\alpha=3$) and long-range ($\alpha=1$) tunneling.
(a) $N=5$ sites profile with flux \(\eta_{\alpha=3} = 0.0194\) and \(\eta_{\alpha=1} = 0.0263\).
(b) $N=6$ sites profile with flux \(\eta_{\alpha=3} = 0.0162\) and  \(\eta_{\alpha=1} = 0.0255\).
(c)-(d) Steady-state populations corresponding to structures in (a)-(b).
(e)-(f) Absolute values of the steady-state density matrix elements  (diagonal removed) for the $N=5$ optimized structures in (a). 
(g)-(h) Absolute values of the steady-state density matrix elements  (diagonal removed) for the $N=6$ optimized structures in (b). 
Other parameters are \(J_{max} = 0.2\) and  $\gamma_l=0.1$.}
    \label{fig:figure17}
\end{figure*}

\begin{figure}[h!]
    \centering
    \includegraphics[width=0.9\linewidth]{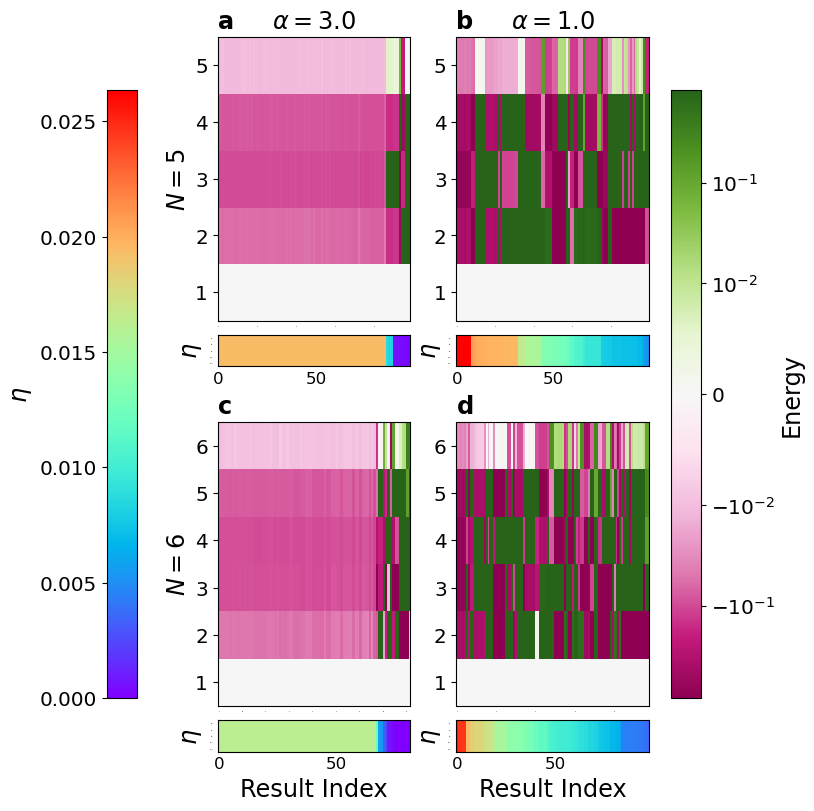}
\caption{Converged optimal energy profiles and corresponding population fluxes found when optimizing transport without environmental interactions. Each column in any panel is a single profile, and the leftmost columns are those plotted in Fig. \ref{fig:figure17}(a)-(b).
Parameters are the same as in Fig. \ref{fig:figure17}.
    }
\label{fig:figure18}
\end{figure}

In this Appendix, we provide complementary results to those in the main text, focusing on chains with $N=5$ and $N=6$ sites. 
When transport is optimized without environmental influences, the resulting optimal energy profiles for five- and six-site systems take on nearly symmetric 'U'-shaped forms, as shown in Fig. \ref{fig:figure17}(a) and (b) for $N=5$ and $N=6$, respectively. 
We observe the following features: (i) All site energies after the first site are negative, with the final site nearly resonant with the initial one.  These trends are consistent with those seen in the three-site model. (ii) No significant even–odd effects are detected; the optimal energy profiles for the five-site and six-site chains are remarkably similar. (iii) The optimal profiles differ substantially between short-range ($\alpha=3$) and long-range ($\alpha=1$) tunneling. In the short-range case, site energies remain relatively close to resonance with the entrance and exit sites. In contrast, the long-range model exhibits significantly larger energy detuning, almost an order of magnitude greater, with central site energies becoming the largest energy scale in the system.

Fig. \ref{fig:figure18} presents the best energy profiles under the coherent model, demonstrating that many local maxima exist for the long-range tunneling ($\alpha=1$)  case.


Turning now to environmental interactions, simulation results for five and six sites under OQS Model I are presented in Fig.
\ref{fig:figure19}.
These results closely resemble those observed for the nine- and ten-site chains in Fig. \ref{fig:figure7}. The energy of the last (exit) site is nearly equal to that of the initial site in both short-range and long-range coupling regimes, while the intermediate site energies sit below. In the short-range case, the energy values are significantly smaller in magnitude compared to the long-range case, and exhibit a monotonic increase from site 2 to the final site. In contrast, long-range energy profiles display a distinct ``wavy” structure between sites 3 and \(N-1\). In all cases, the largest energy gap appears between sites 1 and 2. In correspondence, the steady-state populations and coherences of the five and six-site systems, presented in Fig. \ref{fig:figure19}(c)-(d) and Fig. \ref{fig:figure19}(e)-(h), resemble those shown in Figs. \ref{fig:figure7}(c)-(d) and \ref{fig:figure7}(e)-(h) for nine and ten sites. Remarkably and somewhat deceptively, the optimal energy profiles under coherent conditions, Fig. \ref{fig:figure17}, and  under environmental effects, Fig. \ref{fig:figure19} appear similar. Despite this similarity, differences become evident when examining population distributions, which reveal distinct underlying dynamics: while the population distribution is nearly constant for the coherent model, under local dephasing it exhibits a linear decay.

Fig. \ref{fig:figure20} shows that the optimizer consistently converges to the same solution when starting from different initial conditions, with only mild variations across runs.

Considering now OQS Model II, 
Fig. \ref{fig:figure21} presents optimal energy profiles for five- and six-site chains. Consistent with findings for longer chains in Fig. \ref{fig:figure10}, the optimal structures exhibit ramp-like profiles that support directional flow (and hinder backflow) from site 1 to site $N$.
Checking the robustness of results, Fig. \ref{fig:figure22}  presents the top converged solutions, demonstrating the consistent success of the optimization algorithm.

\begin{widetext}

\begin{figure}[H]
\includegraphics[width=0.9\linewidth]{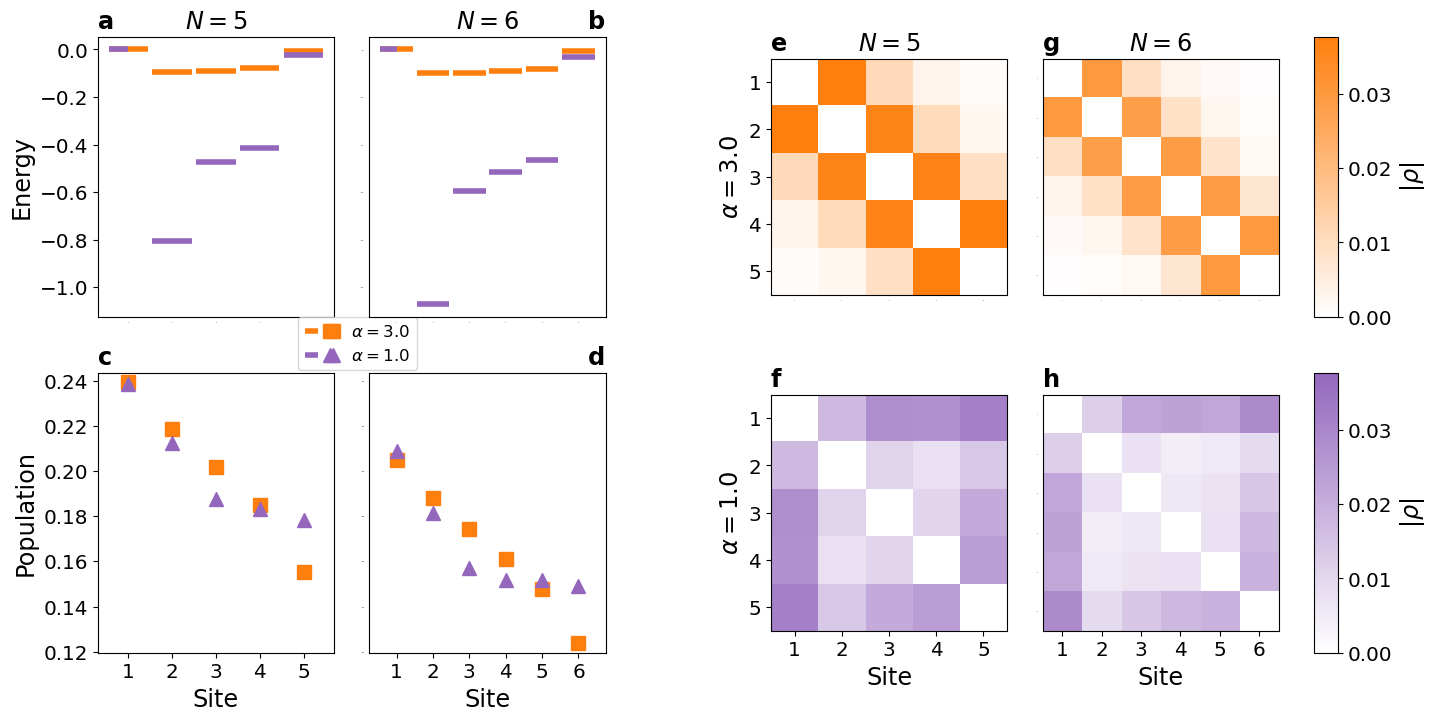}
\caption{OQS Model I: optimized energy landscapes in (a) five-site and (b) six-site chains under the OQS Model I described by Eq. \eqref{lindpd} with \(\Gamma = 0.1\), shown for both short-range ($\alpha=3$) and long-range ($\alpha=1$) couplings.
(a) $N=5$ optimal structure with \(\eta_{\alpha=3} = 0.0156\) and \(\eta_{\alpha=1} = 0.0178\).
(b) $N=6$ optimal structure with \(\eta_{\alpha=3} = 0.0124\) and  \(\eta_{\alpha=1} = 0.0149\).
(c)-(d) Steady-state populations of the energy profiles corresponding to (a)-(b). 
(e)-(f) Absolute values of the steady-state density matrix elements  (diagonal removed) for the $N=5$ optimized structures in (a).
(g)-(h) Absolute values of the steady-state density matrix elements  (diagonal removed) for the $N=6$ optimized structures in (b).
Other parameters are \(J_{max} = 0.2\), and  $\gamma_l=0.1$.}
\label{fig:figure19}
\end{figure}
\vfill
\end{widetext}


\begin{figure}[H]
    \centering
\includegraphics[width=0.9\linewidth]{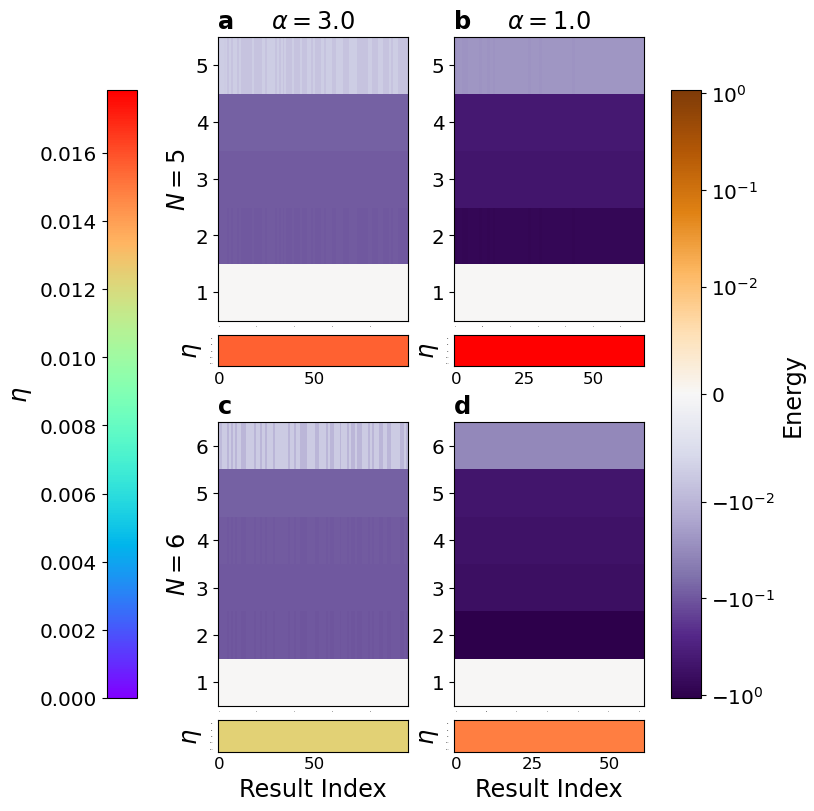}
\caption{
Converged optimal energy profiles and corresponding population fluxes found when optimizing transport in OQS Model I. Each column in any panel is a single profile, and the leftmost columns are those plotted in Fig. \ref{fig:figure19}(a)-(b).
Parameters are the same as in Fig.  \ref{fig:figure19}. }

\label{fig:figure20}
\end{figure}

\begin{widetext}

\begin{figure}[H]
    \centering
\includegraphics[width=0.9\linewidth]{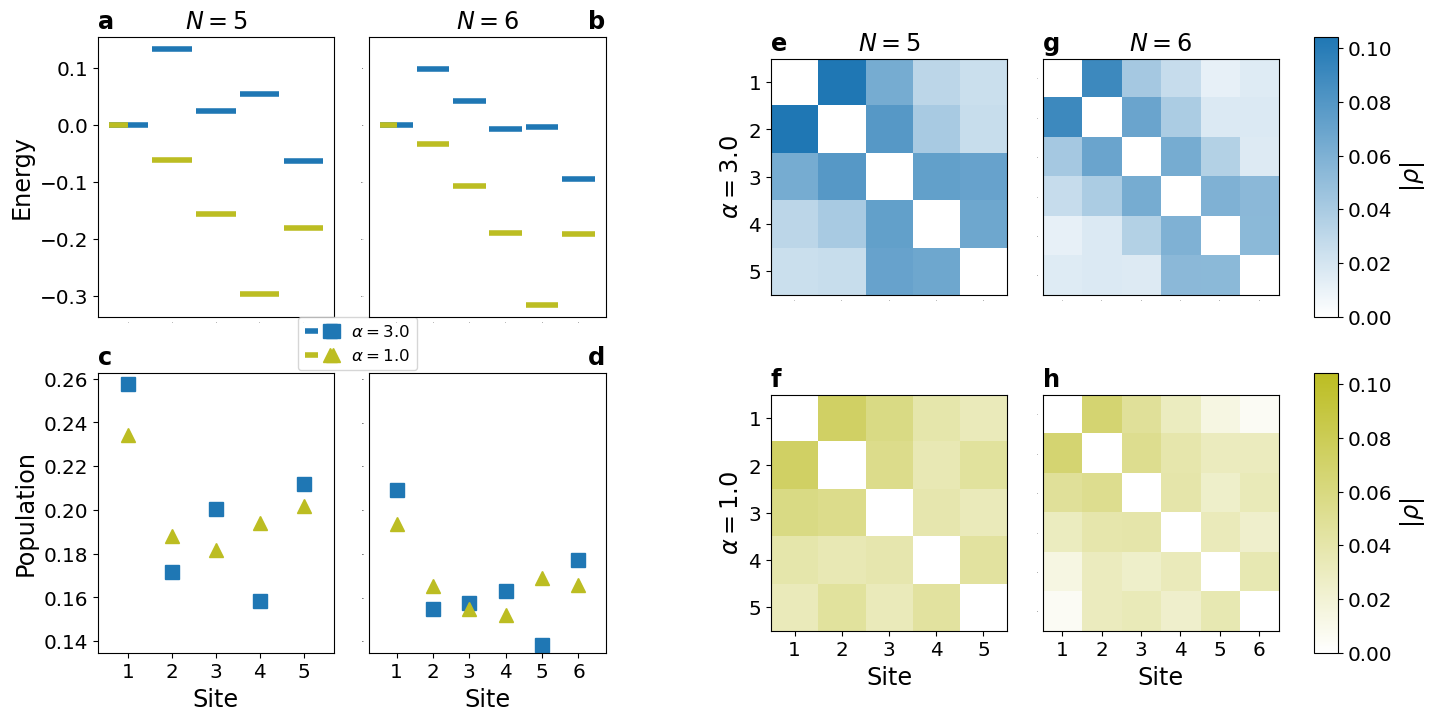}
\caption{OQS Model II: optimized energy landscape in (a) five-site and (b) six-site systems, under the finite-temperature OQS Model II described by Eq. \eqref{lindT} with \(\Gamma_0 = 0.1\) and \(T = 0.2\).
(a) $N = 5$ optimal  profile with $\eta_{\alpha=3} = 0.0212$ and  $\eta_{\alpha=1} = 0.0202$. 
(b) $N = 6$ optimal  profile with $\eta_{\alpha=3} = 0.0177$ and  $\eta_{\alpha=1} = 0.0166$. 
Parameters are \(J_{max} = 0.2\), \(\Gamma_0 = 0.1\), \(\gamma_l = 0.1\), \(T = 0.2\).
(c)-(d) Steady-state population of the energy profiles (a)-(b).
(e)-(f) Absolute values of the steady-state density matrix elements  (diagonal removed) for the optimized structures $N=5$ in (a).
(g)-(h) Absolute values of the steady-state density matrix elements  (diagonal removed) for the optimized structures $N=6$ in (b).
Other parameters are \(J_{max} = 0.2\), \(T = 0.2\), \(\Gamma_0 = 0.1\), and  $\gamma_l=0.1$.}
\label{fig:figure21}
\end{figure}
\vfill
\end{widetext}

\begin{figure}[H]
\includegraphics[width=0.9\linewidth]{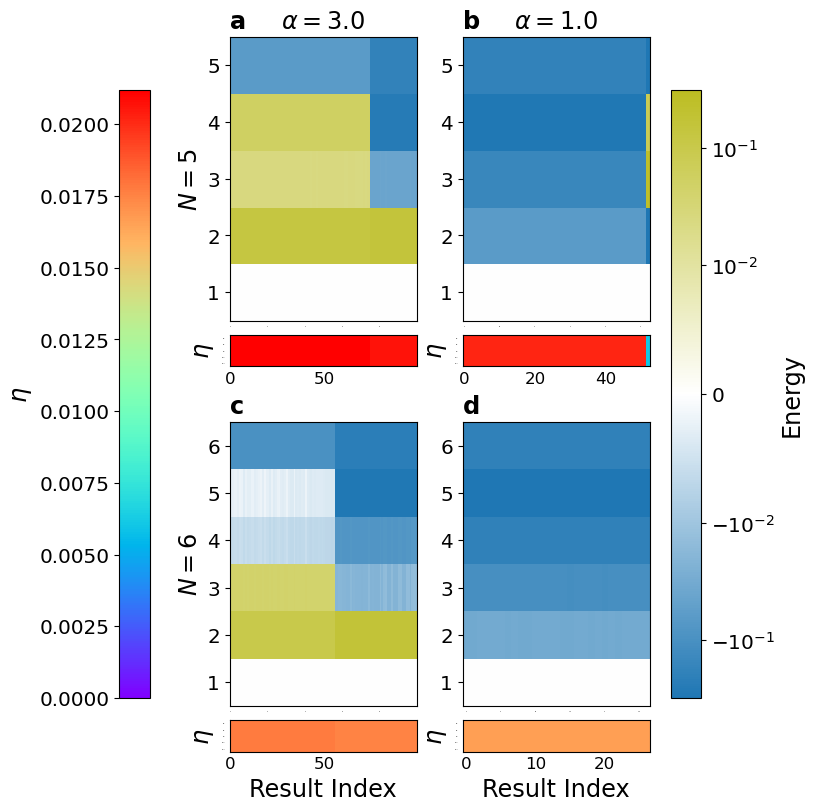}
\caption{Converged optimal energy profiles and corresponding population fluxes found when optimizing transport in OQS Model II. Each column in any panel is a single profile, and the leftmost columns are those plotted in Fig. \ref{fig:figure17}(a)-(b).
Parameters are the same as in Fig.  \ref{fig:figure21}.}
\label{fig:figure22}
\end{figure}

\FloatBarrier
\bibliography{references.bib}

\end{document}